\documentclass[review]{elsarticle}
\usepackage{hyperref}
\pdfoutput=1
\usepackage{etoolbox}
\usepackage{import}
\usepackage{graphicx}
\usepackage[T1]{fontenc}
\usepackage{amsmath}
\usepackage{amssymb}
\usepackage{subcaption}
\usepackage{chngpage}
\usepackage{epsfig}
\usepackage{array}
\usepackage{tabularx}
\usepackage{multirow}
\usepackage{blkarray}
\usepackage{soul}
\usepackage{color}

\usepackage{nomencl}
\makenomenclature
\renewcommand\nomgroup[1]{%
	\item[\bfseries
	\ifstrequal{#1}{A}{Abbreviations}{%
		\ifstrequal{#1}{B}{Latin Symbols}{%
			\ifstrequal{#1}{C}{Greek Symbols}{
				\ifstrequal{#1}{D}{Superscripts}{
					\ifstrequal{#1}{E}{Subscipts}{}}}}}%
]}

\journal{Journal of Sound and Vibration}

\biboptions{sort&compress}

\begin{document}

\begin{frontmatter}

\title{Joint Identification through Hybrid Models Improved by Correlations}

\author{Zeeshan Saeed\corref{mycorrespondingauthor}}
\cortext[mycorrespondingauthor]{Corresponding author}
\ead{zeeshan.saeed@polito.it}
\author{Christian M. Firrone}
\author{Teresa M. Berruti}
\address{Department of Mechanical Engineering, \\ Politecnico di Torino\\
Corso Duca degli Abruzzi 24, 10129 \\
Turin, Italy}


\begin{abstract}
	In mechanical systems coupled with joints, accurate prediction of the joint characteristics is extremely important. Despite years of research, a lot is yet to be learnt about the joints' interface dynamics. 
	The problem becomes even more difficult when the interface Degrees-of-Freedom (DoF) are inaccessible for Frequency Response Function (FRF) measurements. This is, for example, the case of bladed-disk systems with dove-tail or fir-tree type joints. Therefore, an FRF based expansion method called System Equivalent Model Mixing (SEMM) is used to obtain expanded interface dynamics. The method uses numerical and experimental sub-models of each component and their assembly to produce the respective expanded or hybrid sub-models. By applying substructure decoupling to these sub-models, the joint can be identified. However, the joint can be noisy due to expansion and measurement errors which propagate to the hybrid sub-models. 
	
	In this paper, a correlation based approach is proposed in the SEMM method wherein the quality of the expanded sub-models is improved. In this new approach, several expanded models are generated systematically using different combinations of the experimental FRFs and computing a parameter, Frequency Response Assurance Criteria (FRAC), to evaluate quality of the contribution of the different measurements. The lowest correlated channels or FRFs can be filtered out based on a certain threshold value of FRAC. Using the improved hybrid sub-models, the joint identification also shows a remarkable improvement. The test object for the method is an assembly of disk and one blade with a dove-tail joint.
\end{abstract}


\end{frontmatter}


\printnomenclature[1.5cm]
%
%

\section{INTRODUCTION} \label{sec:intro}

Joints are found in many mechanical systems and they tend to influence the system behaviour significantly. From structural dynamics perspective, knowledge of their characteristics is extremely important to accurately predict the structural response. There have been numerous studies in the past that try to identify the joint parameters \cite{Tsai1988,Ren1998,Wang2004,Mehrpouya2013,Meggitt2015, Tol2015,Haeussler2020,Kalaycoglu2018,Latini2020}. Generally, in spectral methods, the joints are identified by inverse substructuring \cite{Zhen2004} or substructure decoupling \cite{DeKlerk2008,Voormeeren2012,DAmbrogio2014}. The inverse substructuring is based on a-priori knowledge of the system to be identified. By using a specific formulation of the dynamic stiffness matrices and measuring only the connection dynamics, the joint properties can be identified. This approach is applied to resilient rubber isolators between two substructures in \cite{Meggitt2015,Haeussler2020}. On the other hand, substructure decoupling methods do not require a-priori knowledge of the system and can be used to decouple any (linear) type of joint. This black-box \cite{Nidhra2012} type identification is more general and allows one to fit or optimize the parameters (stiffness, damping or even mass) later \cite{Tsai1988,Ren1998,Wang2004,Batista2012}. 

The decoupling methods in the class of experimental substructuring have so far been applied to simpler joint interfaces. This is due to difficulties in experimental substructuring \cite{DeKlerk2008} which needs the acquisition of Frequency Response Functions (FRFs) at the interface Degrees-of-Freedom (DoF). Measurement errors of the FRFs at the interface can cause spurious peaks \cite{Rixen2008}. These decoupling methods also often require square FRF matrices and collocated DoF. One of the main challenge then becomes to accurately acquire the drive-point FRFs \cite{Allen2010,Harvie2018}. 

When it comes to complex interfaces such as those found in bladed-disk systems, the conventional methods are not applicable. Their dove-tail or fir-tree interfaces are not reachable for measurements. Therefore, expansion methods are needed. Using the Frequency based Substructuring (FBS) framework \cite{DeKlerk2006}, System Equivalent Model Mixing (SEMM) \cite{Klaassen2018} allows expansion of the measured dynamics on the internal DoF to the interface DoF. 
The peculiarity of SEMM is that it uses different formulations of the same system, coupling the numerical model with experimental measurements performed on a limited number of locations. The final result is a hybrid model mimicking and expanding the dynamics at the measured and unmeasured DoF, respectively.
Its modal domain counterpart System Equivalent Reduction Expansion Process (SEREP) \cite{OCallahan1989} can also be used for expansion. 
However, SEREP requires mode shapes extraction from the measured FRF which is yet another challenging task for light-weight and complex geometries.
Provided that the interface dynamics are observable by internal FRF measurements, SEMM allows to avoid:
\begin{itemize}
	\item direct measurements at the interface; neither for response nor for excitation \cite{Moorhouse2013},
	\item drive point FRF measurements, 
	\item modal parameters estimation especially in regards to the damping.
\end{itemize}
%


This makes SEMM a great candidate for FRF expansion to a complex and inaccessible interface as well as for the identification of the joint properties (linear) by substructure decoupling methods. It should be noted that the decoupling methods are sensitive to errors in measured FRFs \cite{Tsai1988,Voormeeren2012,DAmbrogio2011,Tol2015}. If SEMM is used for the joint identification by the decoupling method, it is imperative that any errors in the hybrid FRFs generated by SEMM are reduced. Since SEMM mixes different models of a structure, discrepancies among them can result in undesirable effects in the hybrid model. The mixed models are, generally, of two types: 
\begin{enumerate}
	\item experimental, which provides compact FRF content but it contains measurement errors including noise.
	\item numerical, which provides a larger DoF set but its accuracy is influenced by material properties, finite element discretization schemes and boundary conditions. 
\end{enumerate} 
It is quite common that some discrepancies arise when the sensor/impact positions in the actual test and the corresponding numerical nodes do not coincide. As a results, inconsistencies propagate in the component hybrid models by SEMM and affect the joint decoupling. In order to reduce the effect of this error as well as the noise in experimental FRFs, a correlation based metric is here introduced in the SEMM method. This new approach will be called \textit{correlated SEMM}. In particular, the correlated SEMM generates multiple hybrid models by systematically using subsets of the experimental model. A statistical correlation parameter called Frequency Response Assurance Criteria (FRAC) is computed between an expanded FRF and an experimental one kept for validation. The procedure is repeated for all the response and input channels. In this way, the highest and lowest correlated channels (DoF) can be identified and the lowest ones filtered out, if needed. 
This approach of computing correlations of the FRFs is analogous to VIKING (Variability Improvement of Key Inaccurate Node Groups) \cite{Thibault2012, Nicgorski2010} which works in the modal domain by improving the modal basis expanded by SEREP through modal correlations.

Moreover, it is good to underline that the constraints or boundary conditions are always difficult to implement \cite{Ewins1995, Smith2016} both in an experimental and in a numerical model. This causes an additional error in the hybrid models. In a recent work of these authors related to the present study on the joint identification \cite{Saeed2020a}, the constraint modelling in one of the substructures -- the disk -- proved to greatly affect the the joint identification result. 


In this paper, the new correlated SEMM method is applied for the identification of the same blade-root joint of the bladed-disk tested in Ref. \cite{Saeed2020a}. 
Both the blade and disk are modelled and tested in free boundary conditions, thereby, minimizing the modelling discrepancies arising from the geometric constraints. The hybrid models' quality is upgraded by using the correlated SEMM method by filtering the uncorrelated FRFs (measured and numerical). As a result, the effect of improved hybrid models is investigated on the joint decoupling. 

The paper is organized as follows: Section~\ref{sec:fbs} briefly covers the FBS method which allows coupling and decoupling of substructures.  Section~\ref{sec:comp_models} presents in detail different DoF classification and the models that are mixed together in the standard SEMM method. Section~\ref{sec:filtering} introduces and elaborates theoretical basis of the correlated SEMM method. The standard and correlated SEMM are then applied to the uncoupled blade and disk test-cases in Section~\ref{sec:filtering_application} followed by the expanded interface description in Section~\ref{sec:interface_vp}. The SEMM methodology is extended to the coupled system and applied to the blade-disk joint identification in  Section~\ref{sec:coupled_models} and Section~\ref{sec:coupled_application}, respectively. 

\nomenclature[A]{DoF}{Degree(s) of Freedom}
\nomenclature[A]{FRF}{Frequency Response Function}
\nomenclature[A]{SEMM}{System Equivalent Model Mixing}
\nomenclature[A]{FBS}{Frequency Based Substructuring}
\nomenclature[A]{FRAC}{Frequency Response Assurance Criteria}
\nomenclature[A]{SEREP}{System Equivalent Reduction Expansion Process}
%
%

\section{FREQUENCY BASED SUBSTRUCTURING} \label{sec:fbs}
\begin{figure}[t!]
	\centering
	\includegraphics[scale = 0.35]{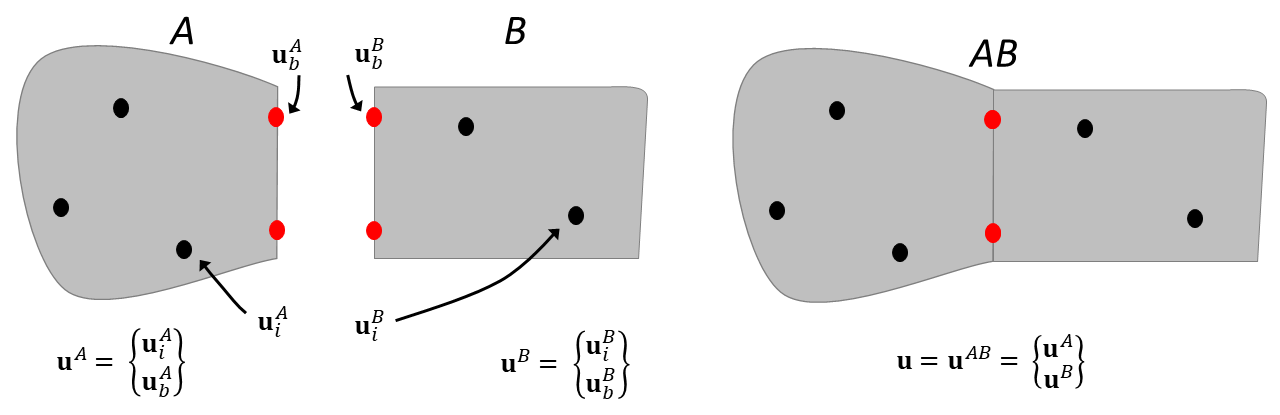}
	\caption{Two dummy substructures $A$ and $B$ with their internal and boundary DoF. They are coupled through the boundary DoF $\mathbf{u}_b$. }
	\label{fig:example_structure}
\end{figure}

This section briefly introduces the key substructuring expressions based on Lagrange Multiplier Frequency Based Substructuring (LM-FBS). Some expressions used here are described in detail in the next sections. Consider two example substructures $A$ and $B$ of Fig.~\ref{fig:example_structure} whose admittances $\mathbf{Y}^A$ and $\mathbf{Y}^B$, respectively, can be computed or measured on the indicated internal $i$ and boundary $b$ DoF. 
%
%
%
The uncoupled receptance (or accelerance) $\mathbf{Y}$ is defined as a block diagonal matrix. 
\begin{equation} \label{eq:FBS_admittance_uncoupled}
	\mathbf{Y} \triangleq diag(\mathbf{Y}^A, \mathbf{Y}^B) = \begin{bmatrix}
		\mathbf{Y}^A & \mathbf{0} \\ \mathbf{0} & \mathbf{Y}^B
	\end{bmatrix}
\end{equation}
\nomenclature[Da]{$A$, $B$}{Substructure identifiers}
\nomenclature[Bb]{$\mathbf{B}$}{Signed Boolean matrix}
\nomenclature[BY]{$\mathbf{Y}$}{FRF matrix, also uncoupled matrix}
The two substructures $A$ and $B$ can be coupled by using the LM-FBS form \cite{DeKlerk2006} as:

%
%
%
%
\begin{equation} \label{eq:FBS_EQM_LMFBS}
	\mathbf{u} = \big(\mathbf{Y} - \mathbf{Y}\mathbf{B}^T ( \mathbf{B} \mathbf{Y} \mathbf{B}^T )^{-1} \mathbf{B} \mathbf{Y} \big) \mathbf{f} \quad \implies \quad  \mathbf{u} = \mathbf{Y}^{AB} \mathbf{f}
\end{equation}
\nomenclature[Bu]{$\mathbf{u}$}{Displacement vector}
\nomenclature[Bf]{$\mathbf{f}$}{External force vector}
where the displacement vector $\mathbf{u}$ consists of all the DoF of $A$ and $B$, as shown in Fig.\ref{fig:example_structure}. The vector of external forces $\mathbf{f}$ is also applied on the same DoF. In the above equation, equilibrium forces on the interface are already eliminated by the use of Lagrange multipliers. The displacement compatibility is applied by the signed Boolean matrix $\mathbf{B}$ such that $\mathbf{B}\mathbf{u} = \mathbf{u}^B_b - \mathbf{u}^A_b = \mathbf{0}$. 
%
Since Eq.~(\ref{eq:FBS_EQM_LMFBS}) will be used several times in this paper, it is represented in the function notation:
\begin{equation} \label{eq:FBS_admittance_LMFBS}
	\mathbf{Y}^{AB} \triangleq fbs(\mathbf{Y},\mathbf{B}) = \mathbf{Y} - \mathbf{Y}\mathbf{B}^T ( \mathbf{B} \mathbf{Y} \mathbf{B}^T )^{-1} \mathbf{B} \mathbf{Y} 
\end{equation}
%
Eq.~(\ref{eq:FBS_admittance_LMFBS}) can also be used: 
\begin{itemize} 
	\item to decouple $A$ from $AB$ to obtain admittance of $B$ by setting $\mathbf{Y} = diag(\mathbf{Y}^{AB}, -\mathbf{Y}^A)$ and calculating $\mathbf{Y}^{B} = fbs(\mathbf{Y},\mathbf{B})$ \cite{Sjovall2008,Voormeeren2012,DAmbrogio2014}. 
	\item to include (linear) effect of the joint flexibility $\mathbf{Y}^J$ by $\mathbf{Y} = diag(\mathbf{Y}^{A}, \mathbf{Y}^J, \mathbf{Y}^B)$ and computing $\mathbf{Y}^{AJB} = fbs(\mathbf{Y},\mathbf{B})$.
	\item to couple and decouple different model descriptions of the same substructure like numerical and experimental expansion purposes, as it will be discussed (although not derived here) in the next section.
\end{itemize}
\nomenclature[Djz]{$J$}{Joint}
Of course, $\mathbf{B}$ has to be appropriately defined in each case. 

\section{SYSTEM EQUIVALENT MODEL MIXING}\label{sec:comp_models}
The System Equivalent Model Mixing (SEMM) method is
an expansion technique that takes
different equivalent FRF models of a structure and couples them so that the dynamics of one are overlaid on the other. It relies on three models, namely, an overlay, a parent and a removed model. The result is a hybrid model that tends to mimic the dynamic behaviour of the structure. In the following subsection, the different DoF classifications and the models that form the basis of SEMM, are described.
%
\subsection{General DoF Description of an FRF Model}
\begin{figure}[t!]
	\centering
	\includegraphics[scale=0.35]{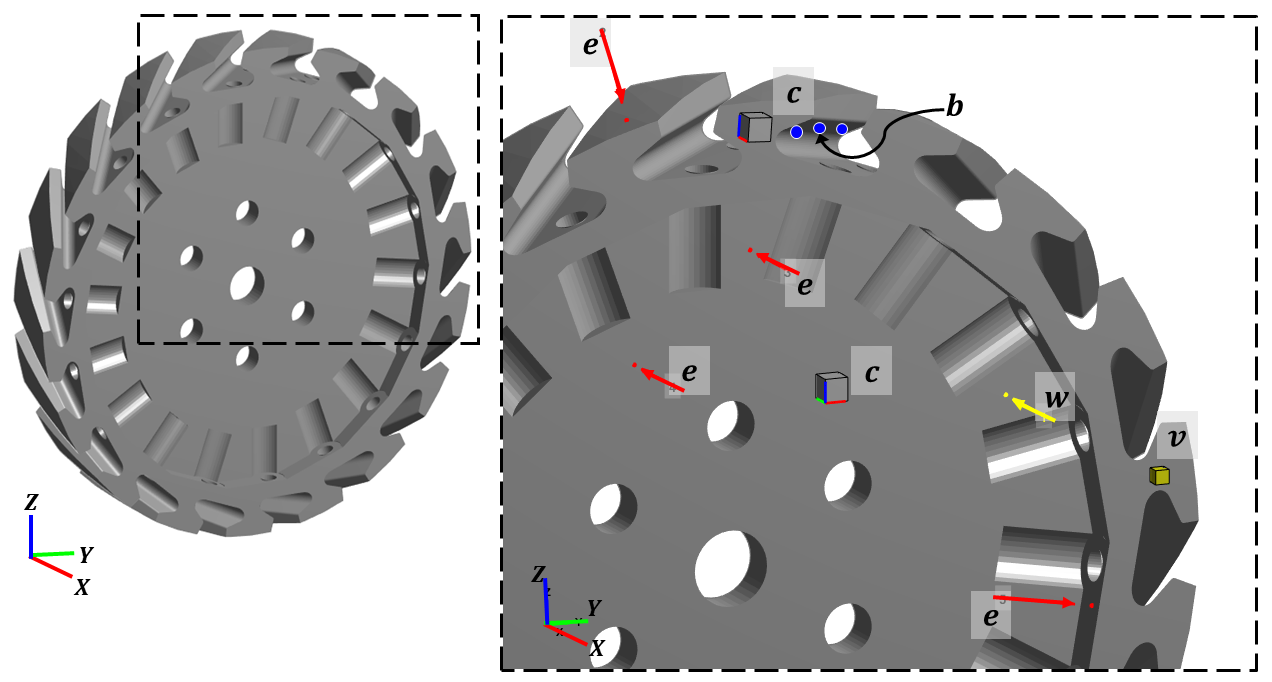}
	\caption{An academic disk on which the different DoF sets are indicated. There are two tri-axial accelerometers labelled $c$, and one uni-axial accelerometer labelled $v$ for validation. Among five $e$ and $w$ labelled impacts, $w$ is designated as a validation impact. All of them form a set of internal DoF. The boundary DoF $b$ are displayed only for one disk-slot. The lack of space in the slot inhibits any direct measurement.}
	\label{fig:Gen_DoF_structure}
\end{figure}

Consider a generic FRF model of a component which consists of internal $\mathbf{u}_i$ and boundary $\mathbf{u}_b$ displacements. On the same degrees of freedom, a set of input forces can also be defined i.e.\ forces $\mathbf{f}_i$ acting on the internal and $\mathbf{f}_b$ on the boundary DoF. The corresponding FRF matrix $\mathbf{Y}$ consists of all the FRFs between output and input DoF.
\begin{equation} \label{eq:DoF_struct_generic}
	\mathbf{Y} = \begin{bmatrix}
		\mathbf{Y}_{ii} & \mathbf{Y}_{ib} \\
		\mathbf{Y}_{bi} & \mathbf{Y}_{bb} \end{bmatrix},  \quad 
	\mathbf{u} = \begin{Bmatrix}
		\mathbf{u}_i \\ \mathbf{u}_b \end{Bmatrix}, \quad \text{and} \quad
	\mathbf{f} = \begin{Bmatrix}
		\mathbf{f}_i \\ \mathbf{f}_b \end{Bmatrix}
\end{equation}
\nomenclature[Ei]{$i$}{Set of internal DoF}
\nomenclature[Eb]{$b$}{Set of boundary DoF}
Since Eq.~(\ref{eq:DoF_struct_generic}) contains the point and transfer functions among all the input and output DoF, it is called a collocated DoF set. Such DoF set is essential for computing the coupled admittance in Eq.~(\ref{eq:FBS_admittance_LMFBS}).
This could easily be obtained from an analytical or numerical model. In order to check the reliability of the numerical model, an experimental validation is always desired. 
%
However, the number of measurements in the experiment is limited due to inaccessibility of some DoF for either response measurement or excitation or even due to limited number of measuring equipment. 
%
%
For bladed-disk interfaces such as dove-tail (Fig.~\ref{fig:Gen_DoF_structure} and Fig.~\ref{fig:blade_virtual_point}) and fir-tree type joints, the boundary DoF are clearly neither measurable nor excitable and hence $\mathbf{Y}_{bb}$, $\mathbf{Y}_{ib}$ and $\mathbf{Y}_{bi}$ can not be obtained experimentally. 
Only the internal FRF $\mathbf{Y}_{ii}$ can be measured since they are accessible. Moreover, not all FRFs can be measured accurately. Especially, the accurate measurement of the drive-point FRFs is very challenging in practice \cite{Ewins1995,Harvie2018}. 
Therefore, the set of internal DoF $\mathbf{u}_{i}$ is divided into different categories based on whether the DoF is a response measurement, input force or used for validation purpose. 
\begin{equation} \label{eq:DoF_struct_internal}
	\mathbf{u}_i = \begin{Bmatrix}
		\mathbf{u}_{c}^T & \mathbf{u}_{v}^T 
	\end{Bmatrix}^T \quad \text{and} \quad \mathbf{f}_i = \begin{Bmatrix}
		\mathbf{f}_{e}^T & 
		\mathbf{f}_{w}^T  
	\end{Bmatrix}^T
\end{equation}
\nomenclature[Ec]{$c$}{Set of internal DoF where responses are measured}
\nomenclature[Ev]{$v$}{Set of internal DoF where responses are measured and reserved for validation}
\nomenclature[Ee]{$e$}{Set of internal DoF where excitations are applied}
\nomenclature[Ew]{$w$}{Set of internal DoF where excitations are applied and reserved for validation}
%
The different subscripts are explained in the following and in Fig.~\ref{fig:Gen_DoF_structure}:
\begin{itemize}
	\item $c$: set of DoF where responses are measured by triaxial accelerometers (or by other sensor types)
	\item $e$: set of DoF where excitations are applied by a modal impact hammer
	\item $v$: set of DoF where responses are measured as $\mathbf{u}_{c}$ but reserved for validation. 
	\item $w$: set of DoF where excitations are applied as $\mathbf{f}_{e}$ but reserved for validation.
\end{itemize}
%
%
\subsection{Experimental FRF Model} \label{sec:comp_models_exp}

Since measurements are not possible on the boundary DoF (interface of Fig.~\ref{fig:Gen_DoF_structure}), an internal DoF based experimental model of FRFs in Eq.~(\ref{eq:DoF_struct_internal}) can be defined as:
\begin{equation} \label{eq:exp_model}
	\mathbf{u}^ {\text{exp}} = \mathbf{Y}^{\text{exp}}  \mathbf{f}^{\text{exp}} 
	\quad \implies \quad
	\begin{Bmatrix} \mathbf{u}_c \\ \mathbf{u}_e  \end{Bmatrix}^\text{exp} = 
	\begin{bmatrix} \mathbf{Y}_{ce} & \mathbf{Y}_{cw} \\ \mathbf{Y}_{ve} & \mathbf{Y}_{vw} \end{bmatrix} ^\text{exp}
	\begin{Bmatrix} \mathbf{f}_e \\ \mathbf{f}_w \end{Bmatrix}^\text{exp}
\end{equation}
\nomenclature[Dexp]{exp}{Measured or experimental model}
\nomenclature[Dov]{ov}{Overaly model} 
The different subscripts of $\mathbf{u}^{\text{exp}}$ and $\mathbf{f}^{\text{exp}}$ indicate that the FRFs are of the non-collocated type, i.e. there is no drive-point FRF. The matrix in Eq.~(\ref{eq:exp_model}) contains all the measurements including the ones used for validation -- subscripts $(\star)_v$ and $(\star)_w$. 
In SEMM, an experimental FRF model of the structure is overlaid on its numerical model (to be discussed in the next subsection) to expand the measured dynamics on the unmeasured DoF \cite{Klaassen2018}. This experimental model is called the overlay model $\mathbf{Y}^{\text{ov}}$ and it can be obtained by setting it equal to $\mathbf{Y}^{\text{exp}}$ or taking its subset. In our case, the overlay model is always a subset of the experimental model since some measurements are used only for validation purposes and they are not included in the overlay model.
\begin{equation} \label{eq:overlay_model}
	\mathbf{u}^ {\text{ov}} = \mathbf{Y}^{\text{ov}}  \mathbf{f}^{\text{ov}} 
	\quad \text{where} \quad
	\mathbf{Y}^{\text{ov}} \subseteq \mathbf{Y}^{\text{exp}}
\end{equation}
If $\mathbf{Y}^{\text{ov}}$ is a subset of $\mathbf{Y}^{\text{exp}}$, different choices for the measurements to include in $\mathbf{Y}^{\text{ov}}$ would produce different hybrid models. This will be further discussed in Section~\ref{sec:filtering} and \ref{sec:filtering_application}.
\subsection{Numerical FRF Model} \label{sec:comp_models_num}

Unlike an experimental model, an FE numerical model allows all DoF to be available, even those not accessible for experimental tests. The definitions of different DoF in Eq.~(\ref{eq:DoF_struct_internal}) along with the boundary DoF $\mathbf{u}_b$ in Eq.~(\ref{eq:DoF_struct_generic}) can be expressed in a square collocated DoF set. 
Note that in an FE model, the mass, stiffness and damping matrices are expressed as dynamic stiffness which is then inverted to compute receptance (or accelerance) form. The numerical FRF model $\mathbf{Y}^\text{N}$ is then expressed as: 
\begin{equation} \label{eq:numerical_model}
	\mathbf{Y}^{\text{N}} \triangleq \mathbf{Y}^{\text{N}}_{gg} = \begin{bmatrix}
		\mathbf{Y}_{ii} & \mathbf{Y}_{ib} \\
		\mathbf{Y}_{bi} & \mathbf{Y}_{bb} \end{bmatrix}^{\text{N}} = \begin{bmatrix} 
		\mathbf{Y}_{cc} & \mathbf{Y}_{ce} & \mathbf{Y}_{cv} & \mathbf{Y}_{cw} & \mathbf{Y}_{cb} \\
		\mathbf{Y}_{ec} & \mathbf{Y}_{ee} & \mathbf{Y}_{ev} & \mathbf{Y}_{ew} & \mathbf{Y}_{eb} \\
		\mathbf{Y}_{vc} & \mathbf{Y}_{ve} & \mathbf{Y}_{vv} & \mathbf{Y}_{vw} & \mathbf{Y}_{vb} \\
		\mathbf{Y}_{wc} & \mathbf{Y}_{we} & \mathbf{Y}_{wv} & \mathbf{Y}_{ww} & \mathbf{Y}_{wb} \\
		\mathbf{Y}_{bc} & \mathbf{Y}_{be} & \mathbf{Y}_{bv} & \mathbf{Y}_{bw} & \mathbf{Y}_{bb} \\
	\end{bmatrix}^{\text{N}}
\end{equation}
\nomenclature[Dnz]{N}{Numerical model} 
\nomenclature[Eg]{$g$}{Set of global DoF} 
%
The subscript $g = \{i, b\} = \{c, e, v, w, b \}$ denotes the global DoF set.
\subsection{Hybrid FRF Model} \label{sec:comp_models_hybrid}
Using the numerical and experimental FRF models, the hybrid model \cite{Klaassen2018} can be computed from the following single-line expression:
%
\begin{equation} \label{eq:semm}
	\mathbf{Y}^{\text{S}} \triangleq  \mathbf{Y}^{\text{S}}_{gg} \triangleq semm(\mathbf{Y}^{\text{N}}, \mathbf{Y}^{\text{ov}})  = \mathbf{Y}^{\text{N}}_{gg} - \mathbf{Y}^{\text{N}}_{gg} (\mathbf{Y}^{\text{N}}_{cg})^+ \big( \mathbf{Y}^{\text{N}}_{ce} - \mathbf{Y}^{\text{ov}}  \big) (\mathbf{Y}^{\text{N}}_{ge})^+\mathbf{Y}^{\text{N}}_{gg} 
\end{equation}
\nomenclature[Ds]{S}{Hybird or expanded model by SEMM}
where $(\star)^+$ represents the Moore-Penrose pseudo inverse. The equation is derived in \cite{Klaassen2018} from the FBS framework \cite{DeKlerk2008}. 
%
The hybrid model has the following properties:
\begin{enumerate}
	\item Within the same set of DoF in the numerical model $\mathbf{Y}^{\text{N}}$, the hybrid model $\mathbf{Y}^{\text{S}}$ has the same experimental features of $\mathbf{Y}^{\text{ov}}$. 
	\item Any measurement errors including noise in the experimental FRFs contained in $\mathbf{Y}^{\text{ov}}$ are transmitted to the FRFs in the hybrid model $\mathbf{Y}^{\text{S}}$. 
	\item Another source of error in the expansion process is the expansion error expressed as norm of the matrix difference $\epsilon = | \mathbf{Y}^{\text{N}}_{ce} - \mathbf{Y}^{\text{ov}} |$. This, of course, depends on how close the numerical model $\mathbf{Y}^{\text{N}}$ is to the experimental one. Measurement errors but also approximation in the models, like not fully realistic constraint conditions, affect $\epsilon$. 
\end{enumerate}
\nomenclature[Ce]{$\epsilon$}{Expansion error}
%
%

\section{CORRELATION ANALYSIS OF HYBRID AND EXPERIMENTAL MODELS} \label{sec:filtering}
In the previous section, a structure's hybrid or expanded model is obtained by coupling its overlay (experimental) and numerical models. The hybrid model can be significantly affected by the discrepancies between the measurements and numerical model. For instance, the location of sensors on the actual structure and the corresponding DoF in its numerical model may not be exactly coincident, thereby, introducing some variations in the respective FRFs. The same holds for the impact positions and direction. Moreover, the numerical model due to its discretization type, material properties and boundary conditions will always have some differences from its experimental counterpart.
%
In this paper, a statistical metric, Frequency Response Assurance Criteria (FRAC), is used \cite{Heylen1996,Grafe1998} to quantify the discrepancies between the FRFs of the two models in a convenient way. In particular, the correlation of FRFs is computed between hybrid model (instead of the numerical model) and the FRFs from measurements kept only for validation and not included in the hybrid model. A strong correlation is indicated by 1 whilst a no correlation is indicated by 0. The FRAC is defined by:
\begin{equation}
	\label{eq:FRAC_shape}
	\phi_{ij} \triangleq FRAC \big(\mathbf{Y}_{ij}^{\text{S}}(\omega), \mathbf{Y}_{ij}^{\text{exp}}(\omega) \big) =  \frac{|\mathbf{Y}_{ij}^{\text{S}}(\omega) \ \mathbf{Y}_{ij}^{\text{exp*}}(\omega)|^2} {\mathbf{Y}_{ij}^{\text{S}}(\omega) \ \mathbf{Y}_{ij}^{\text{S}*}(\omega) . \ \mathbf{Y}_{ij}^{\text{exp}}(\omega) \ \mathbf{Y}_{ij}^{\text{exp}*}(\omega)} 
\end{equation}
\nomenclature[Co]{$\omega$}{Frequency in Hz}
\nomenclature[Cf]{$\phi$}{FRAC}
\nomenclature[Bij]{i, j, k}{Dummy indices}
where $\mathbf{Y}_{ij}^{\text{S}}(\omega)$ and $\mathbf{Y}_{ij}^{\text{exp}}(\omega)$ $\in$ $\mathbb{C}^{n_{\omega} \times 1}$ for each $i$ and $j$. $n_{\omega}$ is number of spectral points and $(\star )^*$ represents the complex conjugate.
\subsection{Filtering Uncorrelated Channels} 
\begin{figure}[t!]
	\centering
	\begin{subfigure}{0.9\textwidth}
		\includegraphics[scale=0.30]{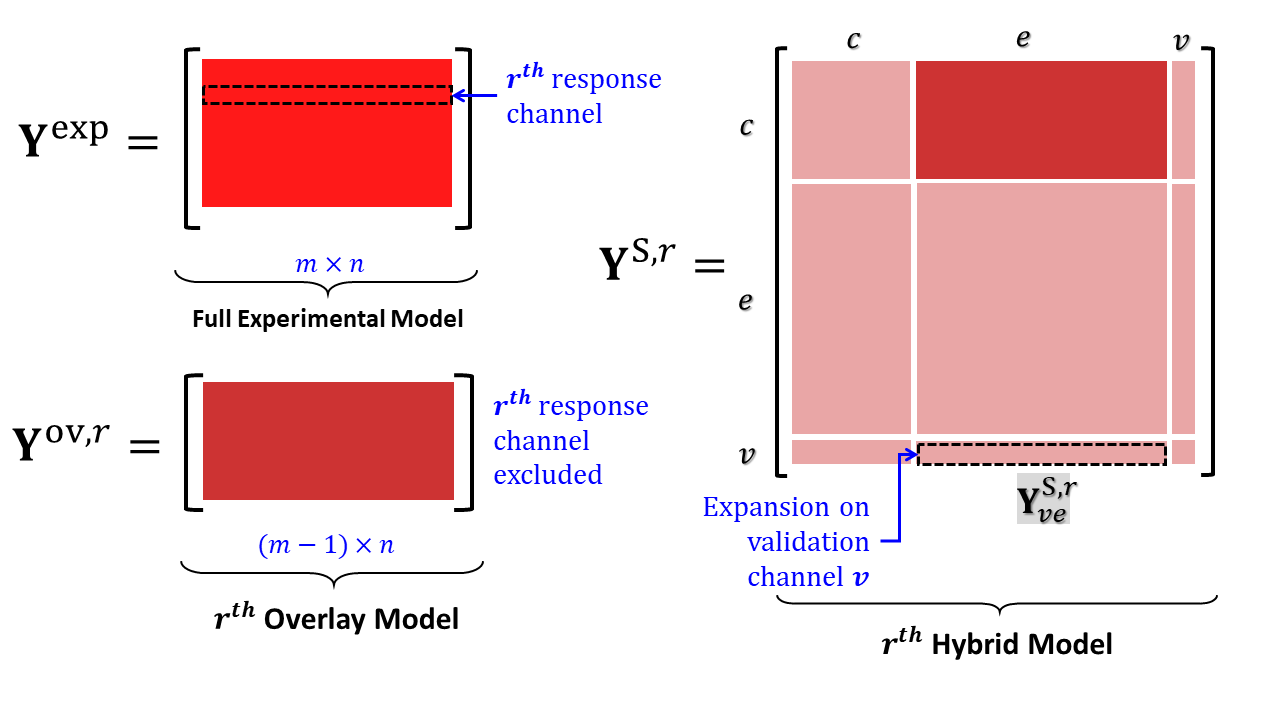}
		\caption{ An example in which the overlay model $\mathbf{Y}^{\text{ov},r}$ is short of the $r^{th}$ response channel (row) in experimental model $\mathbf{Y}^{\text{exp}}$. In the respective $r^{th}$ hybrid model $\mathbf{Y}^{\text{S},r}$, the correlation is calculated between $\mathbf{Y}^{\text{S},r}_{ve}, \mathbf{Y}^{\text{exp}}_{re}$. For simplicity, the DoF set in $\mathbf{Y}^{\text{S},r}$ consists of only $g = \{c, e, v \}$.}
		\label{fig:semm_schematics_rth}	
	\end{subfigure}
	\begin{subfigure}{0.9\textwidth}
		\includegraphics[scale=0.30]{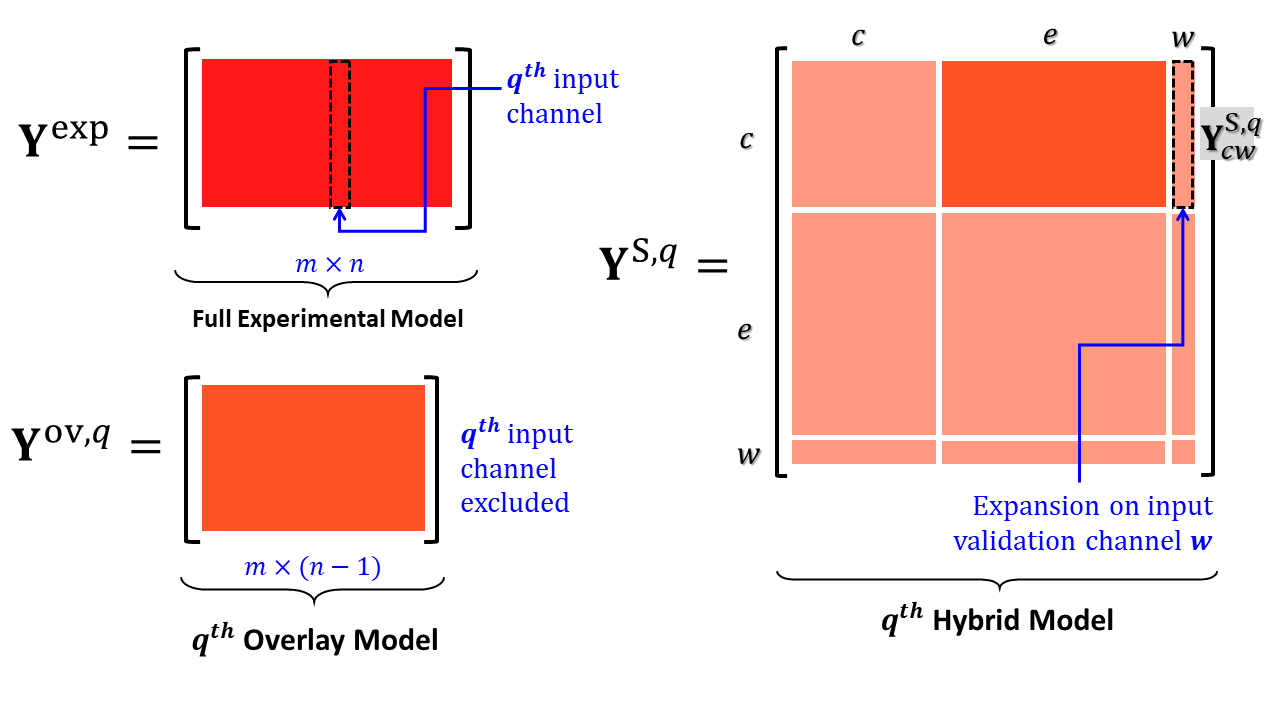}
		\caption{ An example in which the overlay model $\mathbf{Y}^{\text{ov},q}$ is short of the $q^{th}$ input channel (column) in experimental model $\mathbf{Y}^{\text{exp}}$. In the respective $q^{th}$ hybrid model $\mathbf{Y}^{\text{S},q}$, the correlation is calculated between $\mathbf{Y}^{\text{S},q}_{cw}, \mathbf{Y}^{\text{exp}}_{cq}$. For simplicity, the DoF set in $\mathbf{Y}^{\text{S},r}$ consists of only $g = \{c, e, w \}$. }
		\label{fig:semm_schematics_qth}	
	\end{subfigure}
	\caption{Illustration of the different models used to find correlated or uncorrelated response channels and input channels. The DoF set in the hybrid models $\mathbf{Y}^{\text{S}}$ are shown only for the internal DoF. 
		Note the difference in the DoF structure in the top and bottom figure. 
		The colour of $\mathbf{Y}^{\text{exp}}$ is the same in both figures to signify that $\mathbf{Y}^{\text{ov},r}$ and $\mathbf{Y}^{\text{ov},q}$ are its subsets. The same colour appearance of the overlay model in the respective hybrid model shows the mimicking behaviour of those DoF.}
	\label{fig:semm_schematics}
\end{figure}

We introduce the technique of checking the correlation between the FRFs in a systematic way in the SEMM procedure. This new approach, called correlated SEMM, is described in detail in this subsection.
%
The aim is to improve the quality of the substructure hybrid models as much as possible before a subsequent coupled structure model is created (see Section~\ref{sec:coupled_models}).
Therefore, it is proposed that different subsets of experimental model $\mathbf{Y}^{\text{exp}}$, called overlay models, are created to generate the hybrid models by the SEMM method. The process is explained in detail below:
\nomenclature[Br]{$r$}{Index for rows of experimental FRF matrix}
\nomenclature[Bq]{$q$}{Index for columns of experimental FRF matrix}
\nomenclature[Bm]{$m$}{Number of rows of experimental FRF matrix}
\nomenclature[Bn]{$n$}{Number of columns of experimental FRF matrix}
\nomenclature[Davg]{avg}{Average or mean}
\begin{enumerate}
	\item Define an overlay model such that one response channel (a row $\mathbf{Y}^{\text{exp}}_{re}$) from $\mathbf{Y}^{\text{exp}}$ is excluded in the overlay model to be kept for validation, i.e.
	\begin{equation} \label{eq:overlay_rth}
		\begin{aligned}
			\mathbf{Y}^{\text{ov},r} \subset
			\mathbf{Y}^{\text{exp}} : \ \mathbf{Y}^{\text{exp}}_{re} \notin \mathbf{Y}^{\text{ov},r}
		\end{aligned}	
	\end{equation}
	where $r = 1,2,...,m$. Since one channel has been excluded, the size of $\mathbf{Y}^{\text{ov},r}$ is $(m-1) \times n$. The channel $\mathbf{Y}^{\text{exp}}_{re}$ is now considered as the moving validation channel (MVC) and is graphically shown in the upper left part of Fig.~\ref{fig:semm_schematics_rth}.
	\item Perform expansion by the SEMM method with $\mathbf{Y}^{\text{ov},r}$ as per Eq.~(\ref{eq:semm}) to get $\mathbf{Y}^{\text{S},r}$, i.e.\ $\mathbf{Y}^{\text{S},r} = semm(\mathbf{Y}^{\text{N}}, \mathbf{Y}^{\text{ov},r})$.
	%
	%
	\item The corresponding $r^{th}$ expanded channel $\mathbf{Y}^{S,r}_{ve}$ is correlated with $\mathbf{Y}^{\text{exp}}_{re}$ (see Fig.~\ref{fig:semm_schematics_rth}) by computing FRAC, as per Eq.~(\ref{eq:FRAC_shape}). FRAC is computed over a fixed frequency band in Eq.~(\ref{eq:FRAC_shape}) for two given FRFs. However, the explicit dependence of the FRFs on frequency $\omega$ is not shown for the sake of clarity in the above expressions. The FRAC, thus computed for the pairs of FRFs in $\mathbf{Y}^{S,r}_{ve}$ and $\mathbf{Y}^{\text{exp}}_{re}$ are denoted by $\phi_{re}$ and used for calculating $\phi^{\text{avg}}_r$ as follows: 
	%
	%
	%
	\begin{equation} \label{eq:FRAC_avg_r}
		\phi^{\text{avg}}_r = \frac{1}{n} \sum_{j = 1}^{n} \phi_{rj}
	\end{equation}
	The parameter in Eq.~(\ref{eq:FRAC_avg_r}) can be considered an indication of an overall correlation level of the response channel $r$. 
	\item The process is repeated for all the remaining channels up to $r = m$, i.e.\ each time one channel $r$ in $\mathbf{Y}^{\text{exp}}$ is excluded from the $r^{th}$ overlay model.
	\item The low correlated \emph{response channels} are identified based on the average correlation in Eq.~(\ref{eq:FRAC_avg_r}).
\end{enumerate}
In a similar way, by successively excluding the columns from the overlay model, the respective correlations can be computed for the input channels. Fig.~\ref{fig:semm_schematics_qth} illustrates the procedure by excluding $q^{th}$ column from $\mathbf{Y}^{\text{exp}}$. The two schemes of computing correlations are listed side by side in Table~\ref{tab:filtering_response_input_channels} for further clarity.
\begin{table}[t!]
	\centering
	\begin{tabularx}{1.2\textwidth}{ 		>{\raggedright\arraybackslash}l
			| >{\centering\arraybackslash}X 
			| >{\centering\arraybackslash}X  }
		\textbf{Action} & \textbf{Response Channel Correlation} & \textbf{Input Channel Correlation} \\ \hline \hline 
		Define overlay models & $\mathbf{Y}^{\text{ov},r} \subset
		\mathbf{Y}^{\text{exp}} : \ \mathbf{Y}^{\text{exp}}_{re} \notin \mathbf{Y}^{\text{ov},r}$ & $\mathbf{Y}^{\text{ov},q} \subset
		\mathbf{Y}^{\text{exp}} : \ \mathbf{Y}^{\text{exp}}_{cq} \notin \mathbf{Y}^{\text{ov},q}$ \\
		& for $r = 1,2,\ldots,m$ & for $q = 1,2,\ldots,n$ \\
		& $size(\mathbf{Y}^{\text{ov},r}) = (m-1) \times n$ & $size(\mathbf{Y}^{\text{ov},q}) = m \times (n-1)$ \\ \hline
		Generate hybrid models &  $\mathbf{Y}^{\text{S},r} = semm(\mathbf{Y}^{\text{N}}, \mathbf{Y}^{\text{ov},r})$ & $\mathbf{Y}^{\text{S},q} = semm(\mathbf{Y}^{\text{N}}, \mathbf{Y}^{\text{ov},q})$ \\ \hline
		Compute correlations & $\phi_{re} = FRAC( \mathbf{Y}^{\text{S},r}_{ve}, \mathbf{Y}^{\text{exp}}_{re} )$ & $\phi_{cq} = FRAC( \mathbf{Y}^{\text{S},q}_{cw}, \mathbf{Y}^{\text{exp}}_{cq} )$\\
		& $\phi^{avg}_r = \frac{1}{n}\sum_{j = 1}^{n} \phi_{rj}$ & $\phi^{avg}_q = \frac{1}{m}\sum_{i = 1}^{m}  \phi_{iq}$\\ \hline
		Plot and decide & $r = \{ 1,m \}$ vs $\phi^{avg}_r$	& $q = \{1,n\}$ vs $\phi^{avg}_q$\\	\hline	
	\end{tabularx}
	\caption{Summarized action steps to generate overlay and hybrid models in order to find correlations among all DoF or channels (both response and input). The dimension of the overlay matrix is different for response or input channels correlations. Note that $size(\mathbf{Y}^{\text{exp}}) = m \times n$. }
	\label{tab:filtering_response_input_channels}
\end{table}
\subsection{Physical Interpretation} \label{physical_interpret}
The correlation analysis can be interpreted physically from the observability and controllability perspective. At the step $r$, when $r^{th}$ channel is excluded from the construction of the hybrid model, there are $m-1$ response channels which try to observe the $r^{th}$ channel through the SEMM expansion. FRAC as a correlation provides a measure or degree of observability. By repeating the process until $r = m$, each response channel has undergone an observability check (performance review) by the rest of the channels in terms of FRAC. Computing the overall performance, for example, by averaged FRAC values in Eq.~(\ref{eq:FRAC_avg_r}), the best or least observed channels can be identified. A similar interpretation holds also for the input channels from the controllability perspective. In short, excluding the input channel $q$, how well $n-1$ input channels could control the $q^{th}$ input channel.

If one or more channels have low correlation at the end of the process, the following could be the probable reasons:
%
%


%
\begin{enumerate}
	\item the DoF associated with the channel(s) did not have significant dynamic contribution in the selected frequency band. Thus, they could not be fully observed or controlled by the other measured channels.
	\item the location of sensors or impacts and the corresponding DoF in the numerical model were not coincident.
\end{enumerate}
\begin{figure}[t!]
	\centering 
	\begin{subfigure}{0.95\textwidth}
		\centering
		\includegraphics[scale = 0.25]{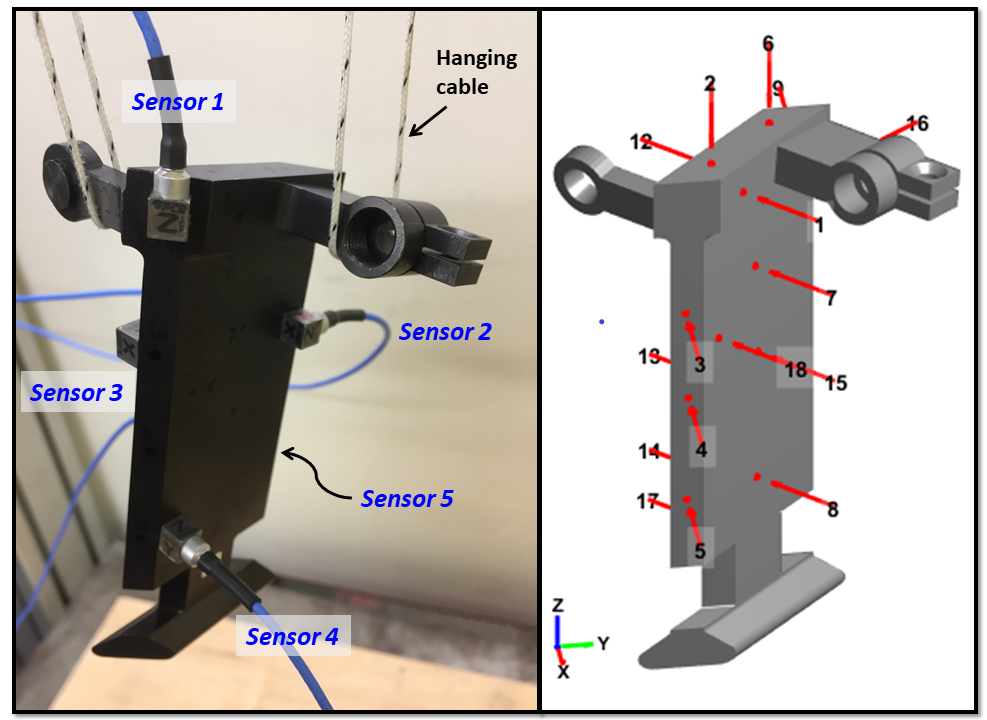}
		\caption{Blade sensor and impact positions. }
		\label{fig:sensors_impacts_blade}
	\end{subfigure}
	\begin{subfigure}{0.95\textwidth}
		\centering
		\includegraphics[scale = 0.3]{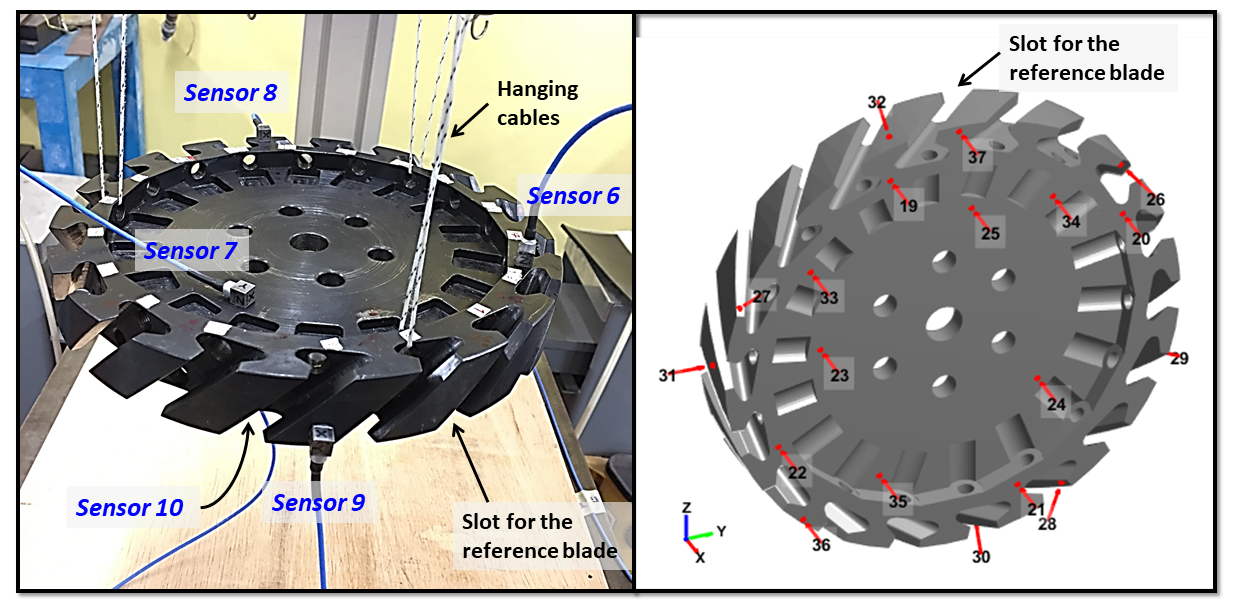}
		\caption{Disk sensor and impact positions.}
		\label{fig:sensors_impacts_disk}
	\end{subfigure}
	\caption{Experimental setup for impact testing for (a) the blade and (b) the disk. The numbering sequence on the disk continues from the blade. Sensor 5 and 10 are not visible for their mounting on the other side. Free boundary condition is realized by supporting the components by flexible wires.}
	\label{fig:sensors_impacts}
\end{figure}
As a result, the least correlated measurement channel(s) can be \emph{filtered out} from $\mathbf{Y}^{\text{exp}}$. In this case, if needed, the corresponding DoF may be kept in the numerical model and thus expanded over in the hybrid model. In another case, the DoF can be \emph{filtered out} from both $\mathbf{Y}^{\text{N}}$ and $\mathbf{Y}^{\text{exp}}$, as it did not contribute much in the global dynamics of the component. 
%
%


Different criteria can be assumed to select the channels to keep after the FRAC analysis. A minimum threshold value criterion would imply that all the channels with a correlation level below the threshold will be disregarded. This could lead to missing the sufficient number of independent measurement channels necessary for an onward analysis. For this reason, it was chosen to define a minimum number $z$ required for the joint identification. In this way, only the $z$ channels with the highest correlation levels are kept, all the other channels are discarded.%
\section{APPLICATION OF CORRELATED SEMM TO A BLADE AND DISK} \label{sec:filtering_application}
In order to find the correlations of hybrid models by the above-mentioned method, two structural components are considered: i) blade and ii) disk (already shown in Fig.~\ref{fig:Gen_DoF_structure}). The effect of the improved hybrid models by the correlation analysis will be studied in Section~\ref{sec:coupled_models} when the blade and disk are assembled.  
Here, the details of the experimental setup and their respective numerical models in the stand-alone configuration are presented.

Two impact testing campaigns were carried out on the blade and disk, as shown in Fig.~\ref{fig:sensors_impacts_blade} and Fig.~\ref{fig:sensors_impacts_disk} along with sensor and impact positions. The sensors are triaxial accelerometers and the excitations are made with a modal impact hammer. The details about the sensors and channels are given in Table~\ref{tab:sensors_channels_details}. 
%
Both the disk and blade were tested while they were hanged with flexible wires. This choice was adopted for two factors:
\begin{enumerate}
	\item The boundary DoF at the root-joint of the blade and disk have to be left unconstrained in the substructuring context. 
	FRFs have to be measured (or expanded by SEMM) on these boundary DoF and hence they have to remain unconstrained. The shroud part on the other end is also left free simply because only one blade is to be coupled to the disk (c.f. Section~\ref{sec:coupled_models}).
	\item The disk was not constrained at its centre to avoid errors due to a non-optimal constraint model \cite{Smith2016}. The impact of the constraint model as a possible source of errors on the final results was shown in a previous work by these authors \cite{Saeed2020a}. 
	%
\end{enumerate}
From these measurements, the accelerance FRFs are collected in $\mathbf{Y}^{\text{exp},A}$ and $\mathbf{Y}^{\text{exp},B}$ for the blade and disk, respectively.
%
%
%
%
\begin{table}[t]
	\centering
	\renewcommand{\arraystretch}{1.1}
	\begin{tabular}{llcc}
		\textbf{Type} & \textbf{Description} & \textbf{Blade} $A$ & \textbf{Disk} $B$ \\ \hline \hline
		\multirow{6}{2cm}{Experimental Setup} & Number of accelerometers & 5 & 5 \\
		& Number of available response channels & 15 & 15 \\
		& Number of useful response channels $(m)$ & 14 & 14 \\
		& Labels for response channels & \{1-8, 10-15\} & \{16-23, 25-30\} \\ 
		& Number of input channels $(n)$ & 18 & 19\\
		& Labels for input channels & \{1--18\} & \{19--37\}\\ \hline
		\multirow{3}{2cm}{Numerical Modelling} & Young's Modulus (GPa)& 190 & 178 \\
		& Density (kg/m$^3$) & 7800  & 7800 \\
		& Fixed interface modes & 200 & 200 \\ \hline
		\multirow{2}{2cm}{Correlation Analysis} & Poorly correlated response channels & ch \# 4 & ch \# 27 \\ 
		& Poorly correlated input channels & ch \# 4, 17 & ch \# 30, 31\\ \hline 
	\end{tabular}
	\caption{Details of experimental and numerical parameters of the blade and disk. The channels with the lowest average FRAC levels are also listed after the correlation analysis. Note a missing channel in the response channel labels. This channel had unusual high noise floor and was not included in the measurements since the beginning of the test campaign. }
	\label{tab:sensors_channels_details}
\end{table}
%

Numerical modelling consisted in creating corresponding FE models from the solid geometries. The discretization was done with Solid elements in ANSYS and with the material properties listed in Table~\ref{tab:sensors_channels_details}. The FE models were then reduced by Hurty-Craig-Bampton \cite{BAMPTON1968} method by retaining only the essential nodes and fixed interface modal amplitudes. The retained DoF corresponded to the nodes of sensors, impacts and the interface. From the reduced systems, accelerance FRFs are computed and stored in $\mathbf{Y}^{\text{N},A}$ and $\mathbf{Y}^{\text{N},B}$ for the blade and disk, respectively.
%

From the above experimental and numerical models, different overlay and hybrid models will be generated in the following sections for the correlation analysis. 
\subsection{Correlation Analysis of the Blade Models}
\begin{figure}[t!]
	\centering
	\begin{subfigure}{0.95\textwidth}
		\centering
		\includegraphics[scale = 0.6]{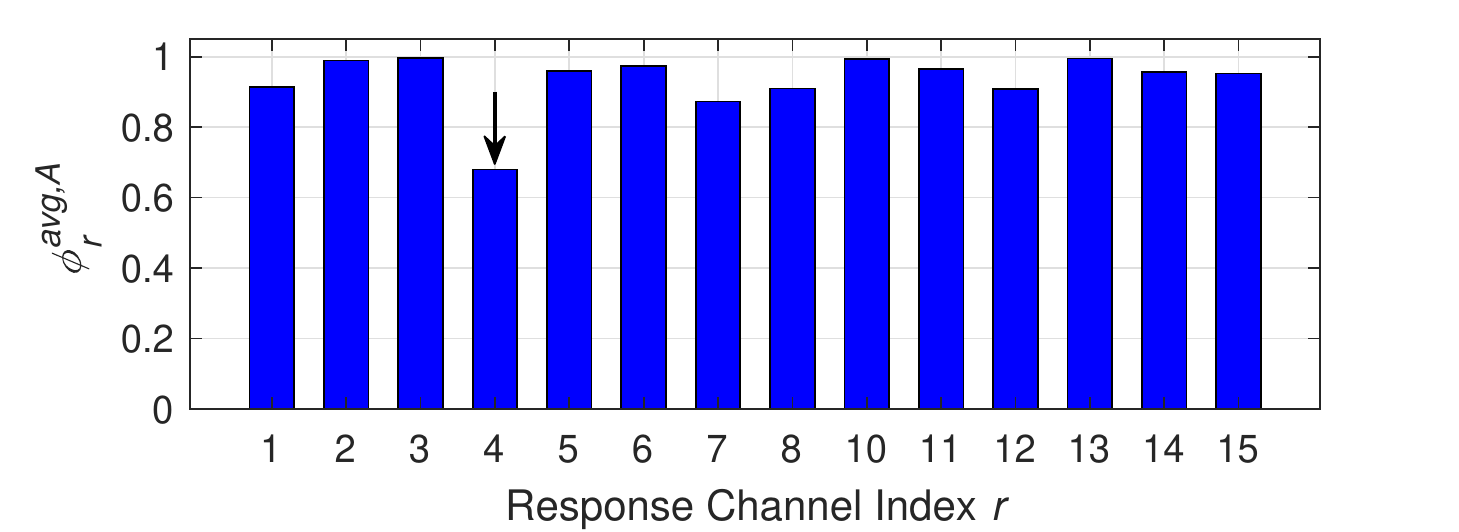}
		\caption{}
		\label{fig:blade_channel_filtering_resp}
	\end{subfigure}
	\begin{subfigure}{0.95\textwidth}
		\centering
		\includegraphics[scale = 0.6]{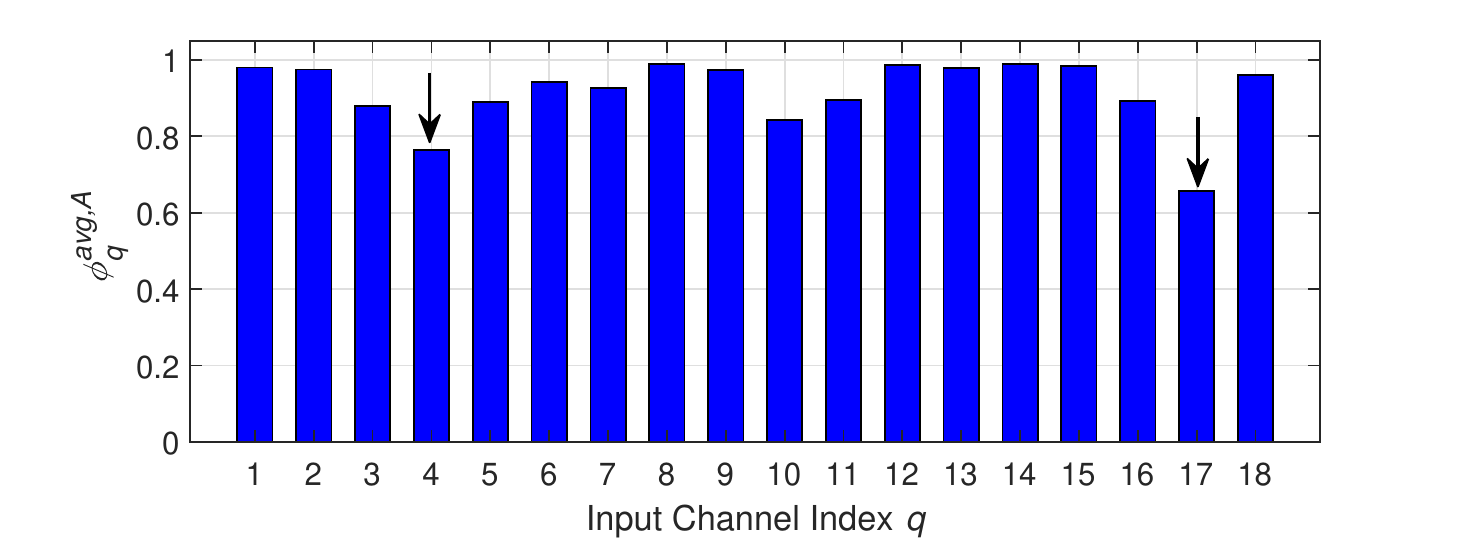}
		\caption{}
		\label{fig:blade_channel_filtering_input}
	\end{subfigure}
	\caption{Average FRAC of the blade $A$ against response channels and (b) input channels. The channels which are removed from measurements based on the lowest FRAC are indicated with arrows.}
	\label{fig:blade_channel_filtering}
\end{figure}

In the first case, the blade's overlay models are generated from the measured FRFs $\mathbf{Y}^{\text{exp},A}$ of size $14 \times 18$ (Table~\ref{tab:sensors_channels_details}). In order to compute correlations for the response channels, $14$ overlay models are generated where each model corresponds to the exclusion of one response channel $r$. The expansion in each hybrid model is checked by computing the FRAC. Since FRAC is computed between two FRFs over a range of frequency, this range is set as $1 - 3000$ Hz. The average FRAC is then plotted against all the $14$ response channels in Fig.~\ref{fig:blade_channel_filtering_resp}. It can be seen that overall correlation levels are higher than $0.80$ except for $r=4$. This means that this channel or DoF could not be \emph{well-observed} by the other channels, when this was removed from $\mathbf{Y}^{\text{ov},A}$. 
\begin{figure}[t!]
	\centering
	\includegraphics[scale = 0.6]{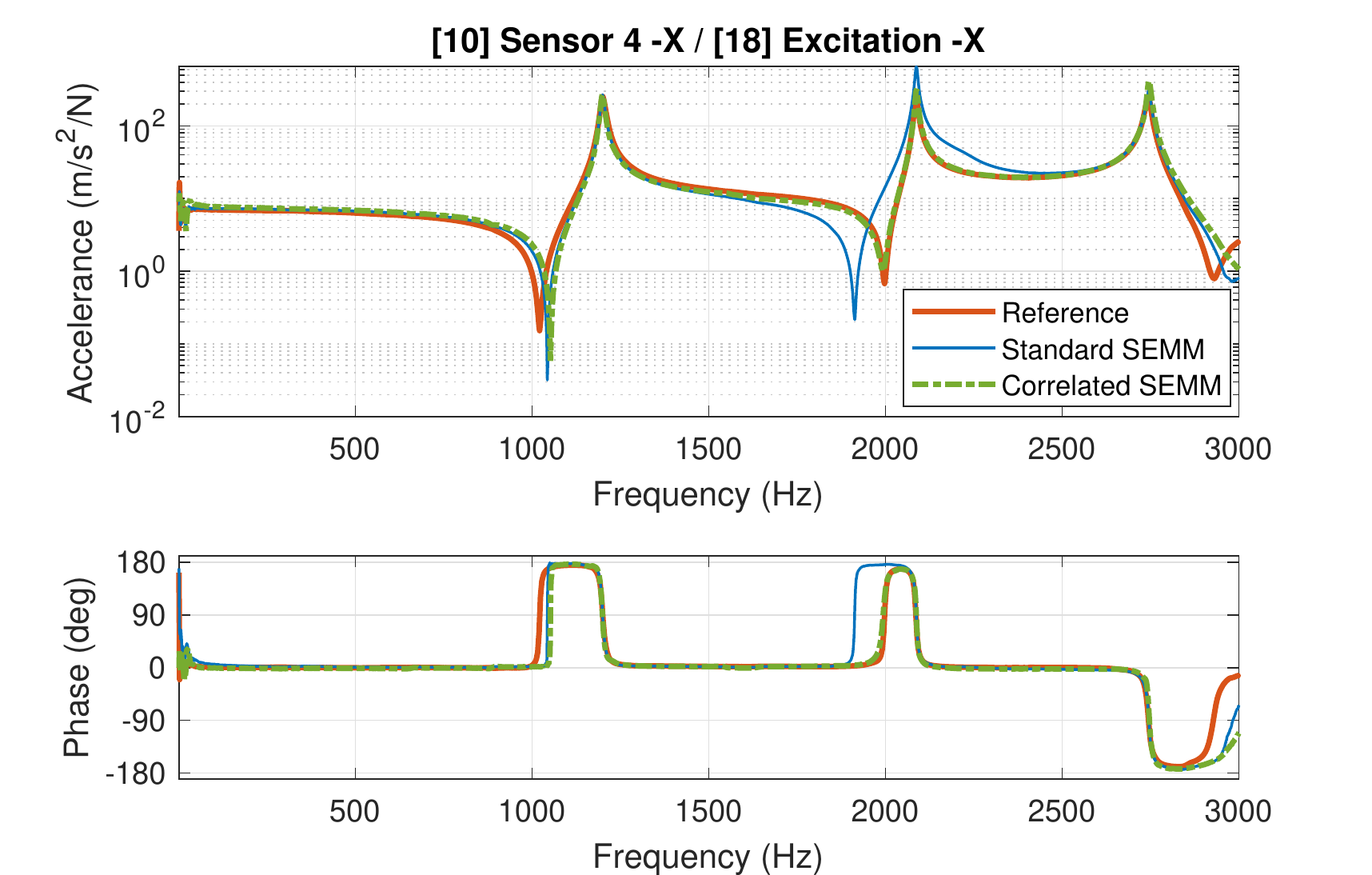}
	\caption{FRFs of the blade with standard SEMM and correlation based SEMM when the lowest correlated channels marked with arrows in Fig.~\ref{fig:blade_channel_filtering} are removed from the measurements. The reference FRF is $\mathbf{Y}^{\text{exp},A}_{10,18}$ with description shown at the top of the FRF.}
	\label{fig:FRF_blade_filtered}
\end{figure} 
%
Following the same method for the input channels by skipping the $q^{th}$ column in $\mathbf{Y}^{\text{exp},A}$ to generate $q^{th}$ overlay and hybrid models, the average FRAC values are plotted as bars in Fig.~\ref{fig:blade_channel_filtering_input} versus the input channels. The correlations are again good for many input channels with the exception of $q=4$ and $q=17$ marked with two arrows. 
%

%
If the channels or DoF with low correlations are retained in the measurements and the standard SEMM method is applied, some of the resulting FRFs may have some inconsistencies. 
In the standard SEMM, all the measured FRFs (except the validation) are included in the overlay model such that $\mathbf{Y}^{\text{ov},A} = \mathbf{Y}^{\text{exp},A}_{ce}$, as per Eq.~(\ref{eq:overlay_model}). In Fig.~\ref{fig:FRF_blade_filtered}, for the sake of validation, an FRF by standard SEMM (thin solid line) is compared with a corresponding experimental FRF (called Reference) and for this reason not included in the construction of $\mathbf{Y}^{\text{S},A}$. 
%
%
%

At the first glance, the standard SEMM method $\mathbf{Y}^{\text{S},A}$ expands the dynamics really well in most of the frequency band. This is because experimental and numerical FRF models are quite close. However, comparing the standard SEMM and reference curves, some inconsistencies are visible especially around $1700-2200$ Hz. In the same figure, it is plotted as dash-dotted line the FRF (labelled: 'Correlated SEMM') obtained from SEMM after filtering out the lowest correlated channels $r=4$, $q=4$ and $q=17$. These channels have been filtered according to the criterion adopted in Section~\ref{physical_interpret}. It can be noticed that this FRF obtained by new correlated hybrid model $\hat{\mathbf{Y}}^{\text{S},A}$ agrees extremely well with the reference FRF both in amplitude and phase (Fig.~\ref{fig:FRF_blade_filtered}). From an a-posteriori check on the measurements of each channel, it came out that the channels discarded by FRAC were not good for different reasons. In detail:
\begin{itemize}
	\item channel $r = 4$ had very low response levels in the shown frequency bandwidth and was, therefore, prone to be easily polluted with noise
	\item input channels $q = 4$ and $q = 17$ did not produce good FRF due to human errors in the impact direction or location.
\end{itemize}

%
\begin{figure}[t!]
	\centering
	\begin{subfigure}{0.95\textwidth}
		\centering
		\includegraphics[scale = 0.6]{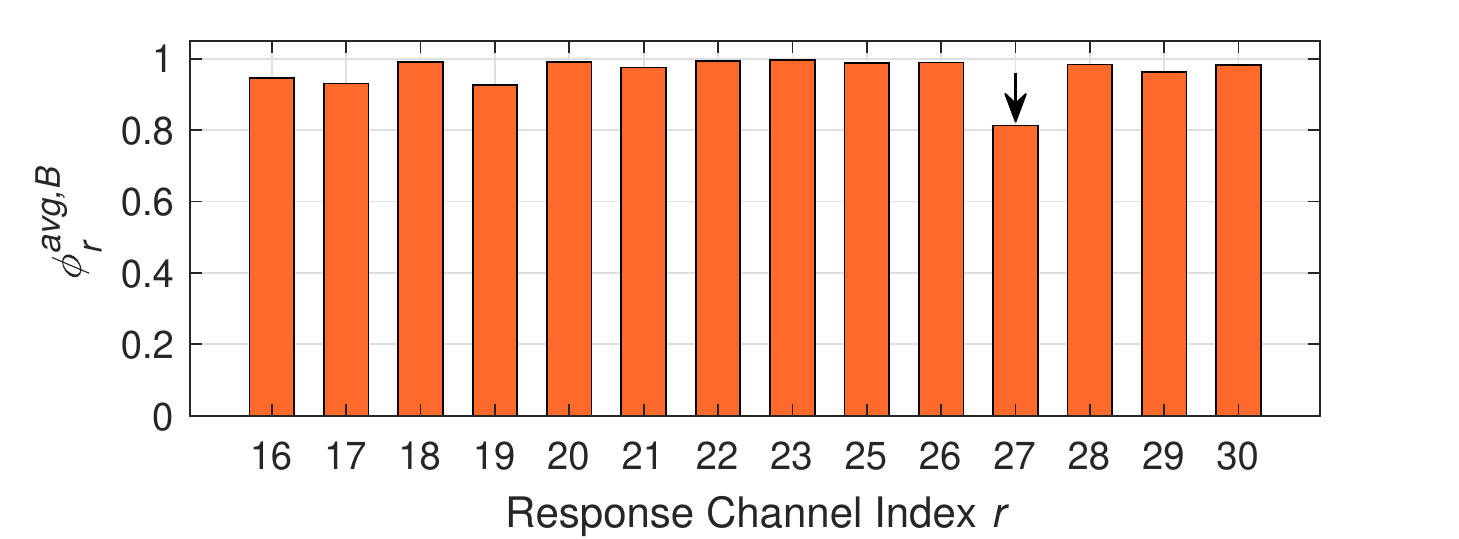}
		\caption{}
		\label{fig:disk_channel_filtering_resp}
	\end{subfigure}
	\begin{subfigure}{0.95\textwidth}
		\centering
		\includegraphics[scale = 0.6]{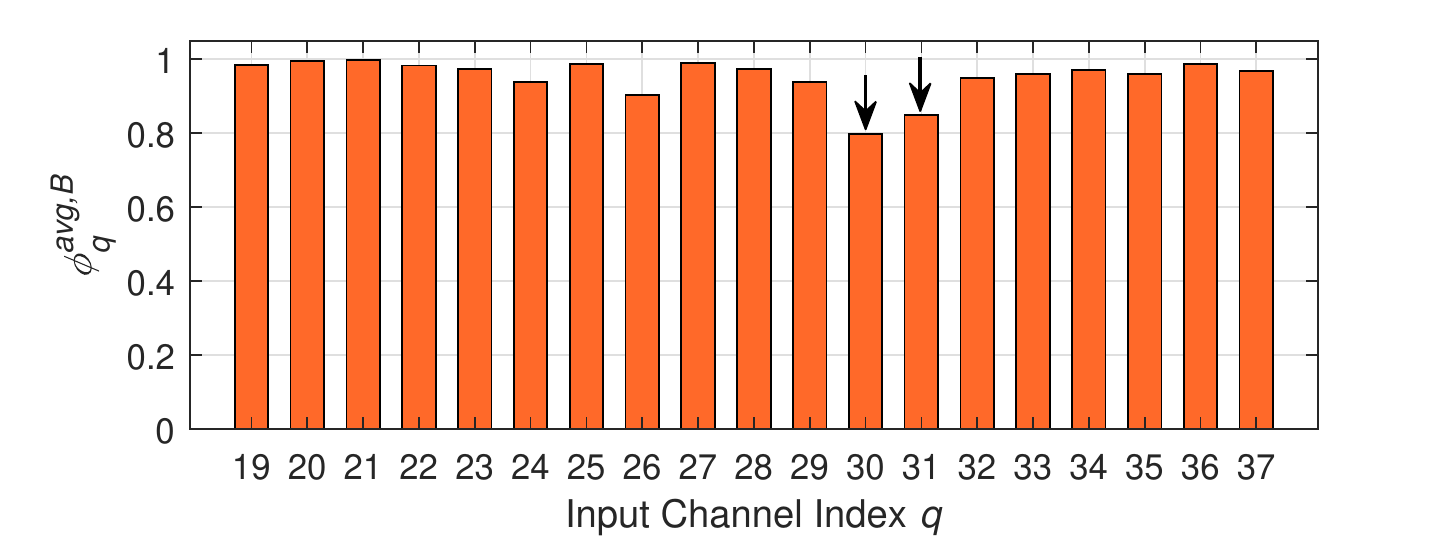}
		\caption{}
		\label{fig:disk_channel_filtering_input}
	\end{subfigure}
	\caption{Average FRAC of the disk $B$ against response channels and (b) input channels. The channels which are removed from measurements based on the lowest FRAC are indicated with arrows.}
	\label{fig:disk_channel_filtering}
\end{figure}

\begin{figure}[t!]
	\centering
	\includegraphics[scale = 0.6]{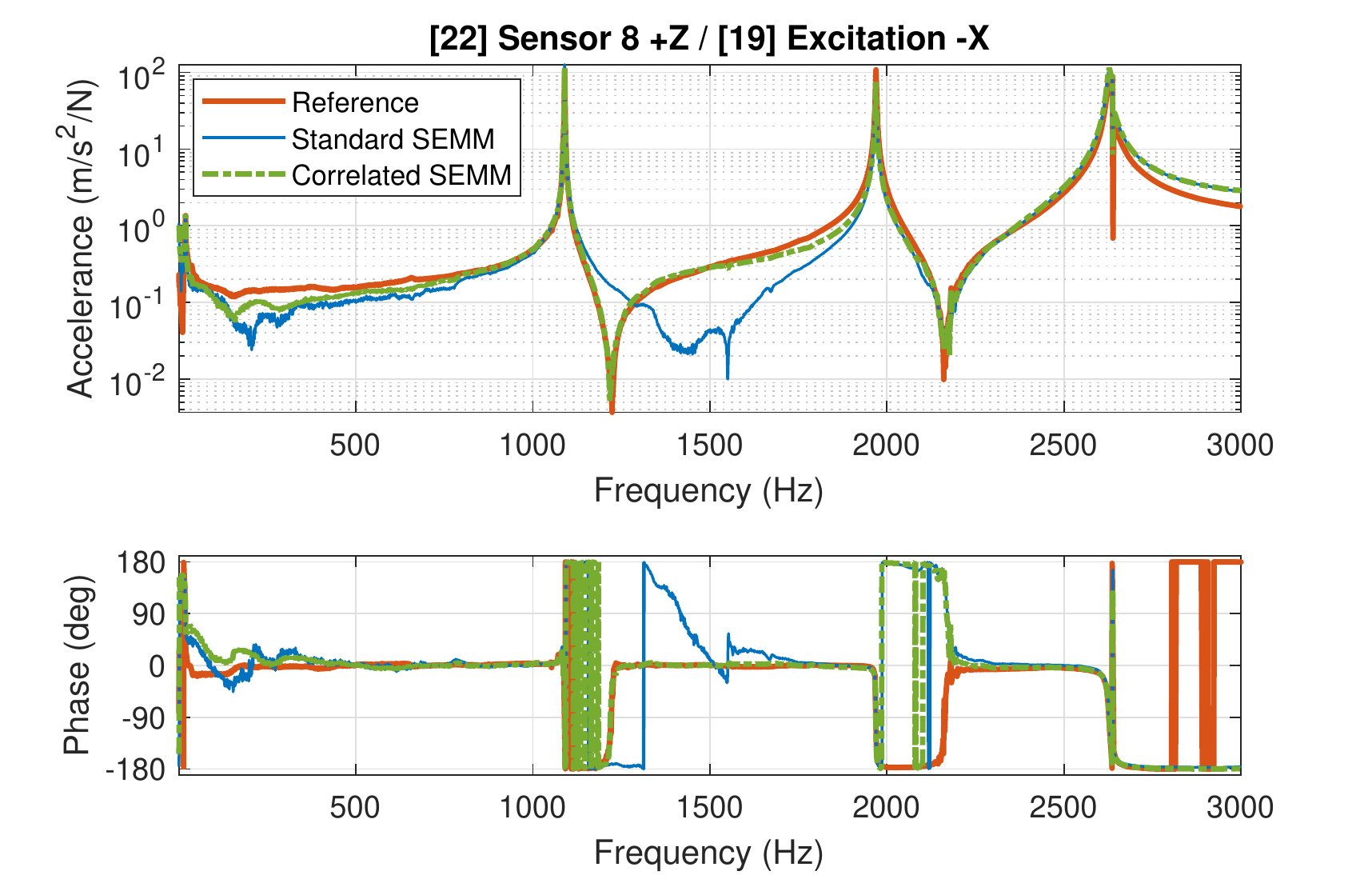}
	\caption{FRFs of the disk with standard SEMM and correlation based SEMM when the lowest correlated channels marked with arrows in Fig.~\ref{fig:disk_channel_filtering} are removed from the measurements. The reference FRF is $\mathbf{Y}^{\text{exp},A}_{22,19}$ with description shown at the top of the FRF.}
	\label{fig:FRF_disk_filtered}
\end{figure}
%
%
%
%
\subsection{Correlation Analysis of the Disk Models}
After the blade's analysis, the disk's correlations are calculated. The disk's FRAC bar graphs similar to that of the blade are shown in Fig.~\ref{fig:disk_channel_filtering} both for the response channels and input channels. One response channel with label $r=27$ and two input channels with label $q=30$ and $q=31$ are found to be the least correlated. By filtering these channels from the experimental (and the overlay) model and regenerating the hybrid model of the disk, the filtering effect is seen in Fig.~\ref{fig:FRF_disk_filtered}. From the figure, it is evident that the standard SEMM using all the measurements produce an FRF that does not overlap with the reference FRF around the first antiresonance from 1200 to 1800 Hz. On the contrary, a remarkably improved FRF is obtained with the correlated SEMM in both the amplitude and phase. 

%
%
%

%
%
\section{INTERFACE DEFINITION} \label{sec:interface_vp}

\begin{figure}[t!]
	\centering
	\includegraphics[scale = 0.4]{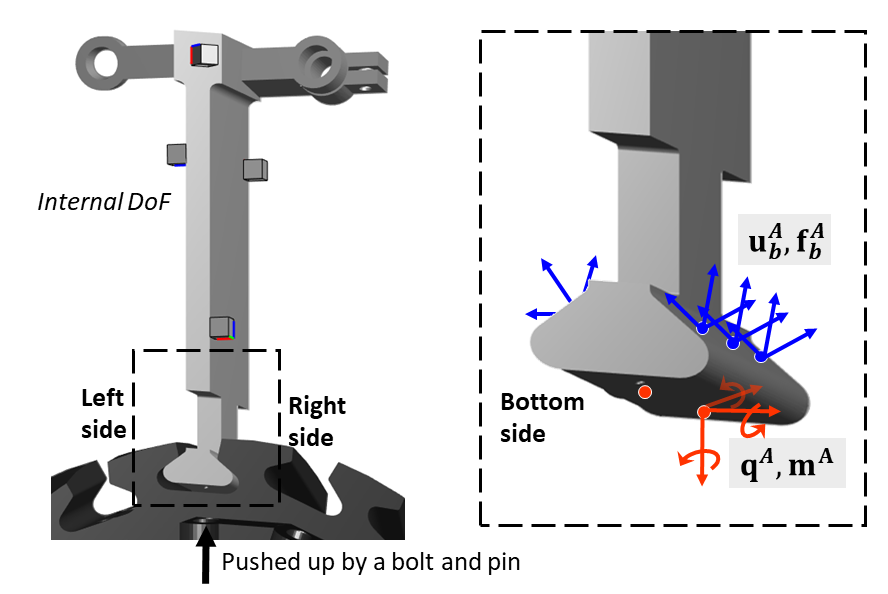}
	\caption{The interface details on the blade. The measured translations in  $\mathbf{Y}^{\text{exp},A}_{ce}$ are expanded to translational boundary DoF $\mathbf{u}^A_{b},\mathbf{f}^A_{b}$. They are then transformed to the indicated two virtual displacements and rotations $\mathbf{q}^A$ and the corresponding virtual forces and moments $\mathbf{m}^A$. For the sake of clarity, virtual translations and rotations are indicated only for one VP.}
	\label{fig:blade_virtual_point}
\end{figure}

The expanded hybrid models in the above sections were related only to the internal DoF in order to be compared and to exclude the low correlated channels. By SEMM, the measured dynamics can be expanded to the inaccessible DoF which are boundary DoF $\mathbf{u}_b$ and $\mathbf{f}_b$. The expanded boundary DoF are shown for the blade in Fig.~\ref{fig:blade_virtual_point} with blue arrows. In an FE model, these DoF are usually the translational DoF of some selected nodes.
Even in the experiments, translations are also easier to measure compared to rotations. However, the interface dynamics are not accurately described if only translational DoF are considered. There have been numerous studies to approximate rotations \cite{DUARTE2000,DAmbrogio2014,DeKlerk,Tol2015} as well as measure them directly \cite{Drozg2018,Cepon2019,Bregar2019}. Based on Equivalent Multi Point Connection (EMPC) \cite{DeKlerk}, a Virtual Point (VP) type interface \cite{VanDerSeijs2014,VanderSeijs2017} can be defined to represent the interface motion by virtual translations and rotations $\mathbf{q}^A$ and virtual forces and moments $\mathbf{m}^A$ (Fig.~\ref{fig:blade_virtual_point}). 
In this work, a VP type interface is considered because the blade-disk joint has three sided-interface i.e.\ left side, right side and bolted pins pushing from the bottom side, as seen in Fig.~\ref{fig:blade_virtual_point}. This means that some uncertainty is associated with the contact at the interface \cite{Saeed2020b} and this will be minimized in a least-squared fashion by the virtual point transformation.
%
%
%
%
%

Consider a hybrid model of the blade $\bar{\mathbf{Y}}^{\text{S},A}$ in which the measured dynamics have been expanded on the boundary DoF $\mathbf{u}_b$, as per the DoF structure of Eq.~(\ref{eq:numerical_model}) and the compact form of Eq.~(\ref{eq:DoF_struct_generic}). The displacements $\mathbf{u}^A$ relate to the virtual point displacements $\mathbf{q}^A$ by:
%
%
%
\begin{equation} \label{eq:vp_u_R_q}
	\begin{Bmatrix} \mathbf{u}^A_i \\ \mathbf{u}^A_b \end{Bmatrix} = 
	\underbrace{\begin{bmatrix} \mathbf{I} & \mathbf{0} \\ \mathbf{0} & \mathbf{R}^A_u \end{bmatrix}}_{\mathbf{R}^A} 
	\begin{Bmatrix} \mathbf{u}^A_i \\ \mathbf{q}^A \end{Bmatrix}
\end{equation}
\nomenclature[A]{VP}{Virtual Point}
\nomenclature[Bqz]{$\mathbf{q}$}{Vector of virtual displacements}
\nomenclature[Bmz]{$\mathbf{m}$}{Vector of virtual forces}
\nomenclature[Brz]{$\mathbf{R}$}{Matrix of coordinates' information about VP}
\nomenclature[Btz]{$\mathbf{T}$}{Matrix of interface displacement or force modes}
where $\mathbf{R}^A_u$ contains the positions and orientations of the DoF in $\mathbf{u}^A_b$ with respect to the virtual point. The vector on the right hand side can then be obtained as
%
%
\begin{equation} \label{eq:vp_q_T_u}
	\begin{Bmatrix} \mathbf{u}^A_i \\ \mathbf{q}^A \end{Bmatrix} = 
	\underbrace{\big( (\mathbf{R}^A)^T \mathbf{R}^A \big)^{-1} (\mathbf{R}^A)^T }_{\mathbf{T}^A}
	\begin{Bmatrix} \mathbf{u}^A_i \\ \mathbf{u}^A_b \end{Bmatrix}
\end{equation}
Since the boundary DoF set in Eq.~(\ref{eq:vp_u_R_q}) is collocated (also see Fig.~\ref{fig:blade_virtual_point}), the same transformation $\mathbf{T}^A$ holds for the virtual forces and moments $\mathbf{m}^A$, i.e.\ $\mathbf{f}^A =  (\mathbf{T}^A)^T \mathbf{m}^A$.
%
%
%
%
%
%
%
Thus, the hybrid FRF matrix $\mathbf{Y}^{\text{S},A}$ for the virtual point interface is computed by:
\begin{equation}
	\mathbf{Y}^{\text{S},A} =  \mathbf{T}^A \; \bar{\mathbf{Y}}^{\text{S},A} \; (\mathbf{T}^A)^T
\end{equation}
Similarly, the boundary DoF in the slots of the disk hybrid models can be transformed to the VP interface. In the discussion to follow, it is assumed that the hybrid models are described by the VP interface. 
\nomenclature[BYz]{$\hat{\mathbf{Y}}$}{FRF matrix of correlated channels}
%
%
\section{COUPLED STRUCTURE MODELS} \label{sec:coupled_models}

In Section~\ref{sec:fbs}, it was discussed how Eq.~(\ref{eq:FBS_admittance_LMFBS}) can be used to decouple the substructures. If the joint can be considered a substructure with its accelerance $\mathbf{Y}^J$, it can be decoupled from its coupled or assembled model $\mathbf{Y}^{AJB}$. In detail, Eq.~(\ref{eq:FBS_admittance_LMFBS}) is to be used as $\mathbf{Y}^J = fbs(\mathbf{Y}, \mathbf{B})$ with $\mathbf{Y} = diag(\mathbf{Y}^{AJB}, -\mathbf{Y}^A, -\mathbf{Y}^B)$. The superscript $J$ emphasizes the explicit presence of the joint or boundary DoF in the coupled model which would not be possible in a directly measured model of the assembly. This is a dual decoupling method \cite{Voormeeren2012}. The methods in \cite{Tsai1988,Tol2015} related to joint identification use primal formulation. In this work, the SEMM method is exploited to generate such a hybrid coupled model $\mathbf{Y}^{\text{S},AJB}$ which explicitly contains boundary dynamics. The method was originally proposed in \cite{Klaassen2019} on a numerical test-case and then applied by the authors of the present paper to the real case in \cite{Saeed2019a, Saeed2020a}. 
The method, which this time uses models generated with the correlated SEMM, is explained here and applied to the blade-disk assembly in Section~\ref{sec:coupled_application}.
\subsection{Coupled Numerical Model}
As well as the individual components, also the assembly needs a hybrid model.
%
%
Since component hybrid models were generated in the preceding sections, they shall be used as a basis for the coupled system's numerical model. In the joint identification context, the boundary dynamics should also be present. Therefore, at the beginning, a guessed joint accelerance $\mathbf{Y}^{J}_k$ is introduced. So, the coupled numerical accelerance $\mathbf{Y}^{\text{N},AJB}$ can be written as:

\begin{equation} \label{eq:coupled_numerical}
	\mathbf{Y}^{\text{N},AJB}_k = fbs(\mathbf{Y},\mathbf{B}) \quad \text{with} \quad 
	\begin{cases}
		\mathbf{Y} = diag( \mathbf{Y}^{\text{S},A}, \mathbf{Y}^{J}_k, \mathbf{Y}^{\text{S},B} ) \\[10pt] 
		\mathbf{B} = 
		\begin{blockarray}{cccccc} 
			\mathbf{u}^A_i & \mathbf{q}^A & \mathbf{q}^{J,A} & \mathbf{q}^{J,B} & \mathbf{u}^B_i & \mathbf{q}^B \\
			\begin{block}{[cc|cc|cc]} 
				\mathbf{0} & -\mathbf{I} & \mathbf{I} & \mathbf{0} & \mathbf{0} & \mathbf{0} \\ 
				\mathbf{0} & \mathbf{0} & \mathbf{0} & \mathbf{I} & \mathbf{0} & -\mathbf{I} \\
		\end{block}
		\end{blockarray} 
	\end{cases}
\end{equation}
%
%
%
where $\mathbf{q}^{J,A}$ and $\mathbf{q}^{J,B}$ represent the DoF of the joint which couple to substructure $A$ and $B$, respectively. The index $k=0,1,2,\ldots$ indicates the iteration number, as the joint is not known a priori. The joint needs to be updated at every iteration and as a results, $\mathbf{Y}^{\text{N},AJB}_k$ would also be updated. The iterative nature of the method has been explained in \cite{Saeed2020a}. 
\begin{figure}[t!]
	\centering
	\includegraphics[scale = 0.35]{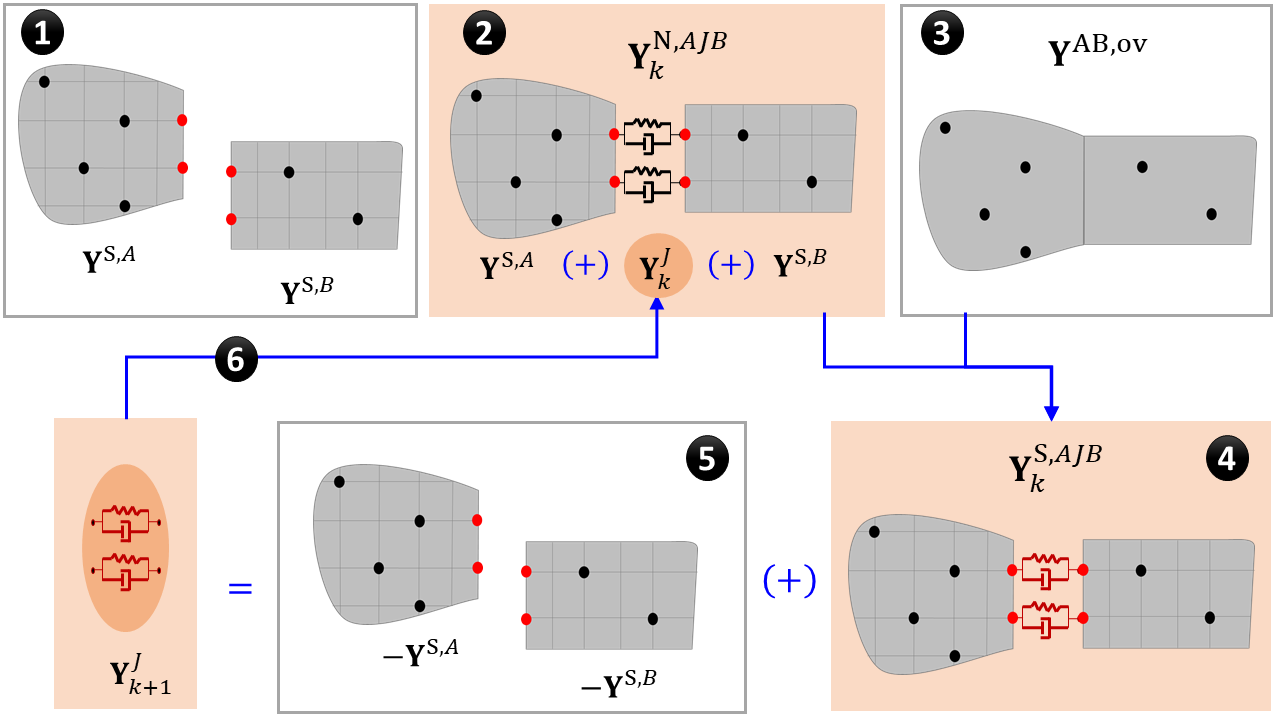}
	\caption{Illustration of the SEMM method applied to an assembled system in order to decouple the joint. The example joint shown is a spring and damper, but the decoupling method is not limited to it. The quantities in coloured blocks (2,4 and 6) are updated at each iteration $k$. The sign $(+)$ indicates coupling of substructures. }
	\label{fig:semm_decoupling_schematics}
\end{figure} 
\subsection{Coupled Experimental Model}
The measured FRFs on the coupled system $\mathbf{Y}^{\text{exp},AB}$ contain joint dynamics implicitly. 
%
From this experimental model, an overlay model $\mathbf{Y}^{\text{ov},AB}$ is taken as a subset and imposed on the numerical model $\mathbf{Y}^{\text{N},AJB}_k$. Some channels in $\mathbf{Y}^{\text{exp},AB}$ are not used for the overlay model and are kept for validation. Note that this overlay model remains the same at each iteration.
\subsection{Coupled Hybrid Model}
With the above coupled numerical and overlay models, the hybrid model is generated from Eq.~(\ref{eq:semm}) such that 
\begin{equation} \label{eq:coupled_semm}
	\mathbf{Y}^{\text{S},AJB}_k = semm( \mathbf{Y}^{\text{N},AJB}_k, \mathbf{Y}^{\text{ov},AB})
\end{equation}
The hybrid model is also updated iteratively. Since $\mathbf{Y}^{\text{N},AJB}_k$ has a guessed linear joint (or no joint) at $k=1$ which may be far from the actual one, so there may exist a high expansion error $|\mathbf{Y}^{\text{N},AJB}_k - \mathbf{Y}^{\text{ov},AB}| $. 

\subsection{Joint Decoupling} \label{sec:joint_decoupling}
With the above different coupled models, it is now possible to decouple the joint dynamics by Eq.~(\ref{eq:coupled_FBS_admittance_LMFBS})
\begin{equation} \label{eq:coupled_FBS_admittance_LMFBS}
	\mathbf{Y}^{J}_{k+1} = fbs(\mathbf{Y},\mathbf{B}) \quad \text{with} \quad \mathbf{Y} = diag(\mathbf{Y}^{\text{S},AJB}_k, -\mathbf{Y}^{\text{S},A}, -\mathbf{Y}^{\text{S},B})
\end{equation}
$\mathbf{Y}^{J}_{k+1}$ is then substituted in Eq.~(\ref{eq:coupled_numerical}) to update the numerical model $\mathbf{Y}^{\text{N},AJB}_{k+1}$, to subsequently generate an updated hybrid model $\mathbf{Y}^{\text{S},AJB}_{k+1}$ in Eq.~(\ref{eq:coupled_semm}) and, thereafter, to decouple the joint $\mathbf{Y}^{J}_{k+2}$. The iterative process is graphically illustrated in Fig.~\ref{fig:semm_decoupling_schematics}. 
At each iteration, the updated joint improves because this is the only part that is updating the numerical model $\mathbf{Y}^{\text{N},AJB}$. 
The process is repeated until the expansion error $\epsilon = |\mathbf{Y}^{\text{N},AJB}_k - \mathbf{Y}^{\text{ov},AB}|$ between the numerical and the experimental model is reduced below a given threshold.
%
%
It should be noted that:
\begin{enumerate}
	\item The initial guess of $\mathbf{Y}^{J}_{k}$ at $k=1$ can be a blank joint i.e. the substructures can be left uncoupled \cite{Klaassen2019, Saeed2020a}.
	\item $\mathbf{Y}^{J}_{k+1}$ obtained by LM-FBS equation has all the rows and columns corresponding to the DoF of both the coupled and uncoupled models. It is necessary to retain only the independent entries \cite{DAmbrogio2011}.
	\item The method converges faster by using weighted pseudo-inverses with higher weights assigned to the boundary DoF in Eq.~(\ref{eq:coupled_semm}). This aspect has been deeply discussed in \cite{Saeed2020a}.
\end{enumerate}

\section{APPLICATION OF SEMM TO THE BLADE-DISK ASSEMBLY} \label{sec:coupled_application}
\begin{figure}[t!]
	\centering
	\begin{subfigure}{0.49\textwidth}
		\centering
		\includegraphics[scale = 0.21]{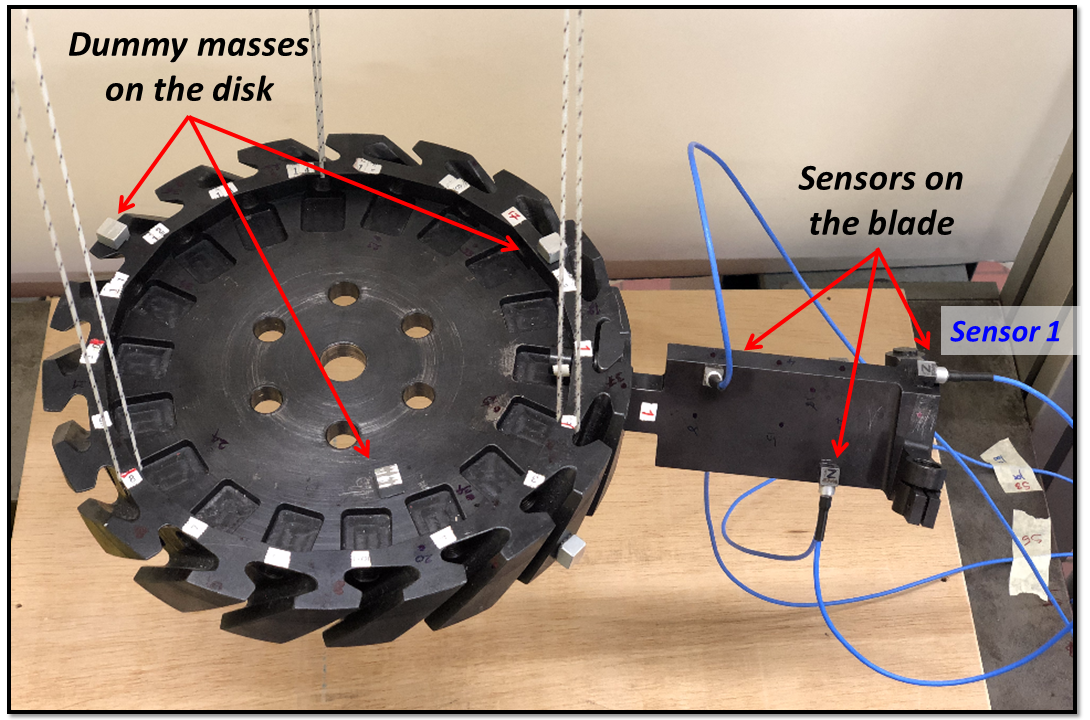}
		\caption{}
		\label{fig:coupled_measurements_sensors_blade}
	\end{subfigure}
	\begin{subfigure}{0.49\textwidth}
		\centering
		\includegraphics[scale = 0.21]{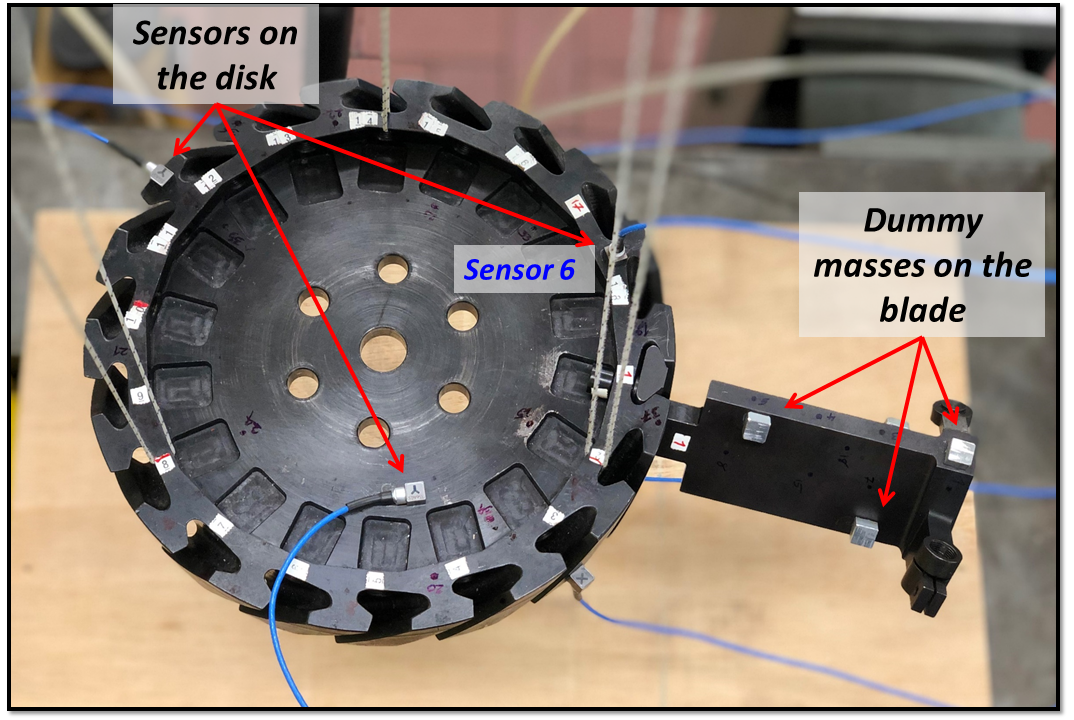}
		\caption{}
		\label{fig:coupled_measurements_sensors_disk}
	\end{subfigure}
	\caption{Experimental setup of the blade coupled to the disk. Due to limited number of sensors and channels in the data acquisition system, the campaign was completed first by (a) mounting the sensors on the blade and the dummy masses on the disk and then by (b) mounting sensors on the disk and the dummy masses on the blade. Each dummy mass value is equivalent to the sensor's nominal mass. The sensor and impact positions were preserved exactly as  Fig.~\ref{fig:sensors_impacts_blade} and \ref{fig:sensors_impacts_disk}, respectively. Positions of sensor 1 and sensor 6 are indicated for reference. }
	\label{fig:couple_measurements}
\end{figure}
In this section, the SEMM method is applied to the assembly of the blade and disk of Section~\ref{sec:filtering_application}. Fig.~\ref{fig:couple_measurements} shows the blade-disk assembly in free constraint conditions. The impact test campaign was carried out on the coupled system with the same sensor and impact positions as in the case of blade and disk alone. Since there were limited number of sensors and data acquisition channels, they had to be mounted once on the blade and then on the disk. To obtain the joint by decoupling, the blade and disk models (substructures) have to be decoupled from the coupled model. The sensor masses are considered part of the substructure models. Therefore, by placing dummy masses on one component while the sensors are on the other, the additional mass effect is cancelled. The set of FRF measurements on the assembly is denoted by $\mathbf{Y}^{\text{exp},AB}$ whose size is $28 \times 37$ (see Table~\ref{tab:sensors_channels_details}). 
From this, the assembly overlay model $\mathbf{Y}^{\text{ov},AB}$ is taken as a subset of size $27 \times 36$, while one response channel and one input channel are left as the validation channels. In detail, the experimental FRF used for validation is $\mathbf{Y}^{\text{exp},AB}_{vw} = \mathbf{Y}^{\text{exp},AB}_{2,27}$.
\subsection{The Decoupled Joint}
\begin{figure}[t!]
	\centering
	\begin{subfigure}{0.46\textwidth}
		\centering
		\includegraphics[scale = 0.5]{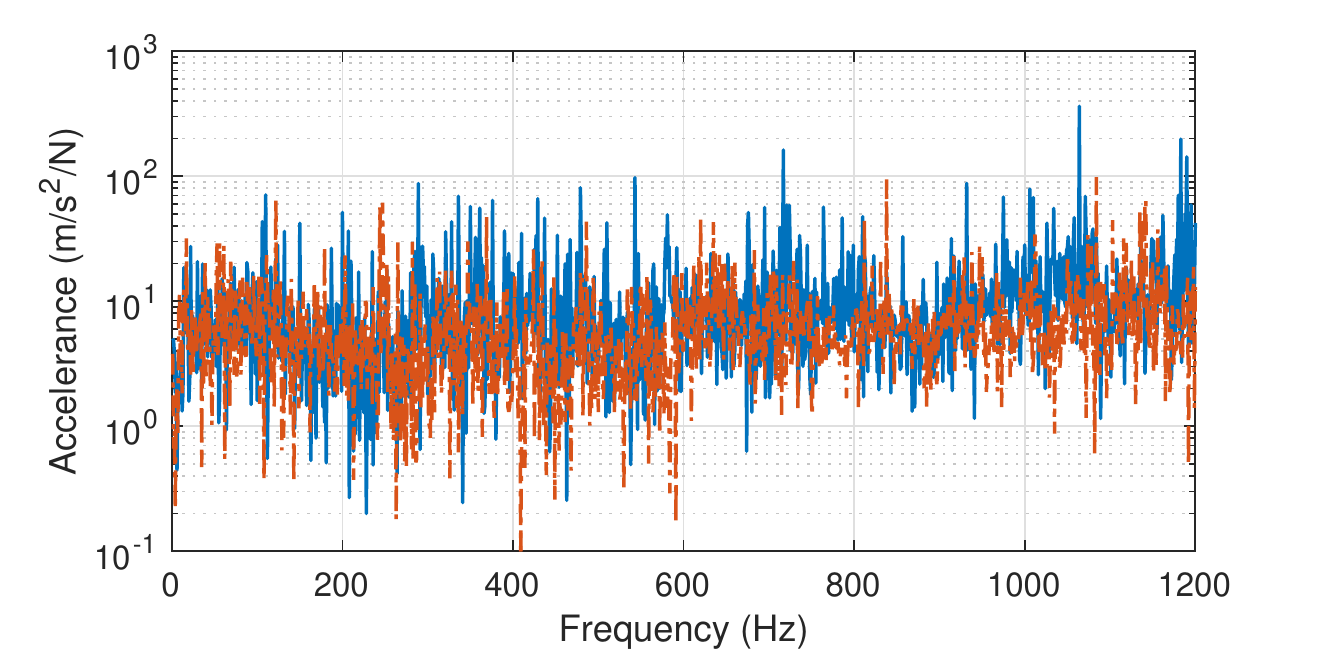}
		\caption{$\mathbf{Y}^{J}_{13,13}$ X/X}
		\label{fig:the_joint_trans_bad}
	\end{subfigure} \qquad
	\begin{subfigure}{0.46\textwidth}
		\centering
		\includegraphics[scale = 0.5]{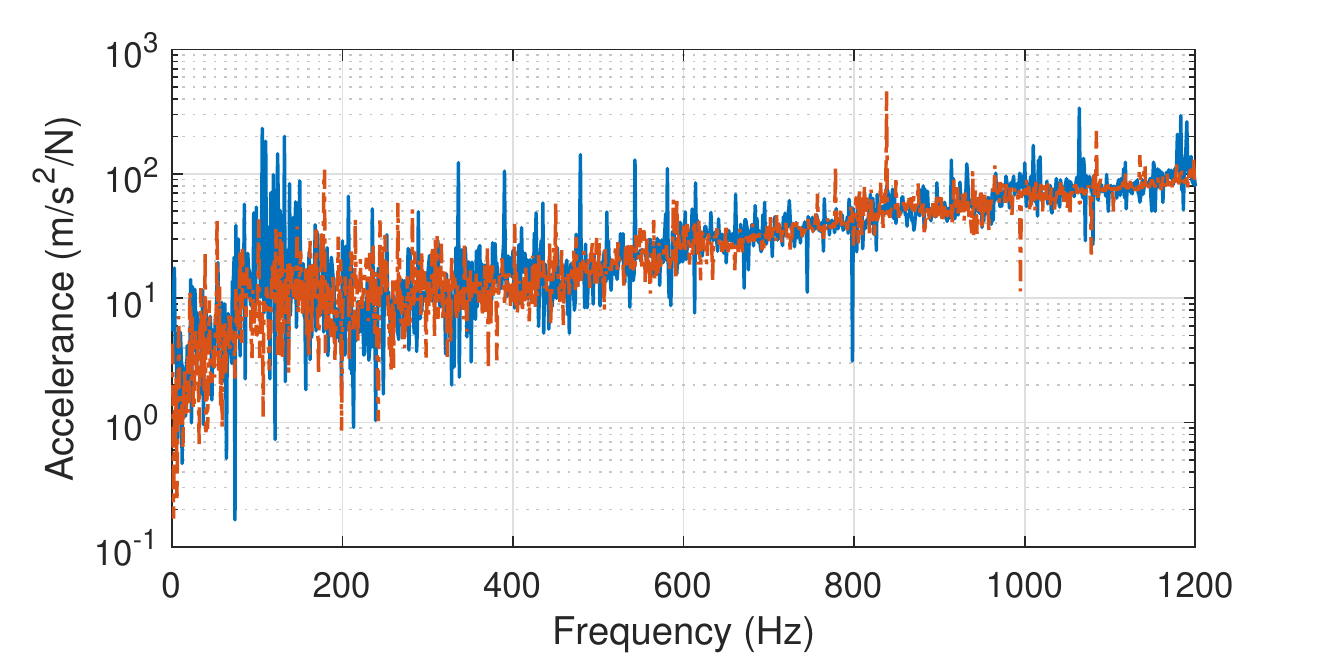}
		\caption{$\mathbf{Y}^{J}_{15,15}$ Z/Z}
		\label{fig:the_joint_trans_good}
	\end{subfigure} \\
	\begin{subfigure}{0.46\textwidth}
		\centering
		\includegraphics[scale = 0.5]{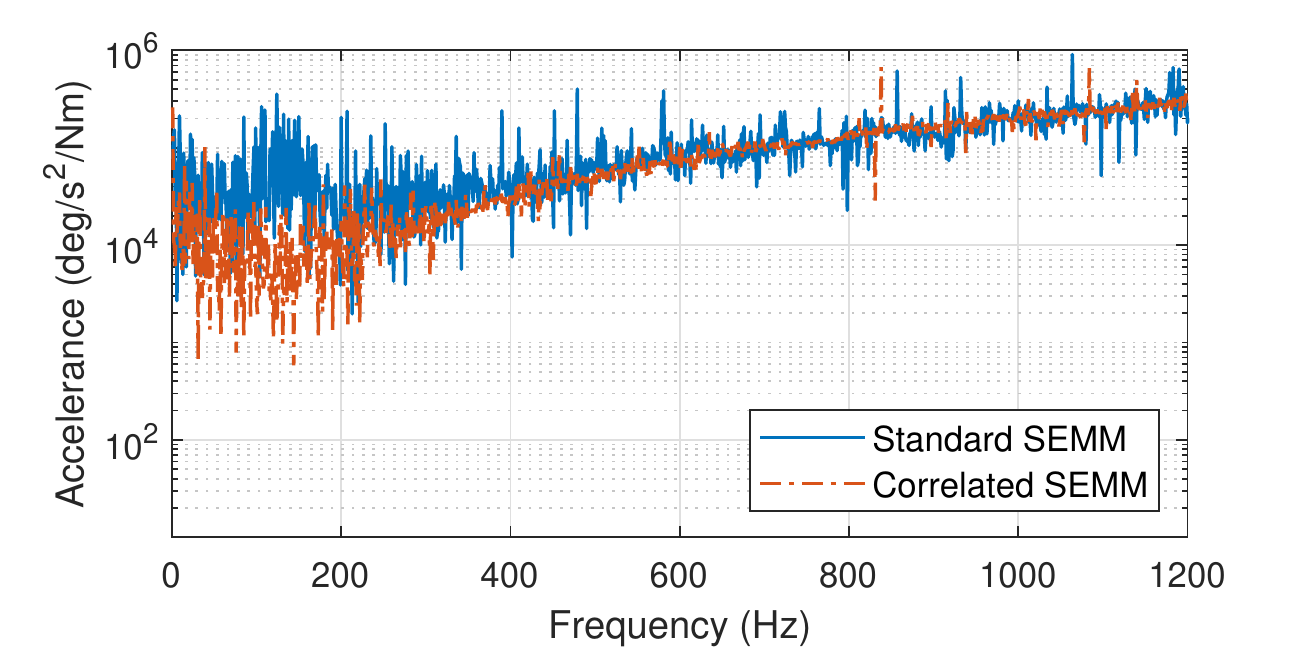}
		\caption{$\mathbf{Y}^{J}_{16,16}$ RX/RX}
		\label{fig:the_joint_rot_bad}
	\end{subfigure} \qquad
	\begin{subfigure}{0.46\textwidth}
		\centering	
		\includegraphics[scale = 0.5]{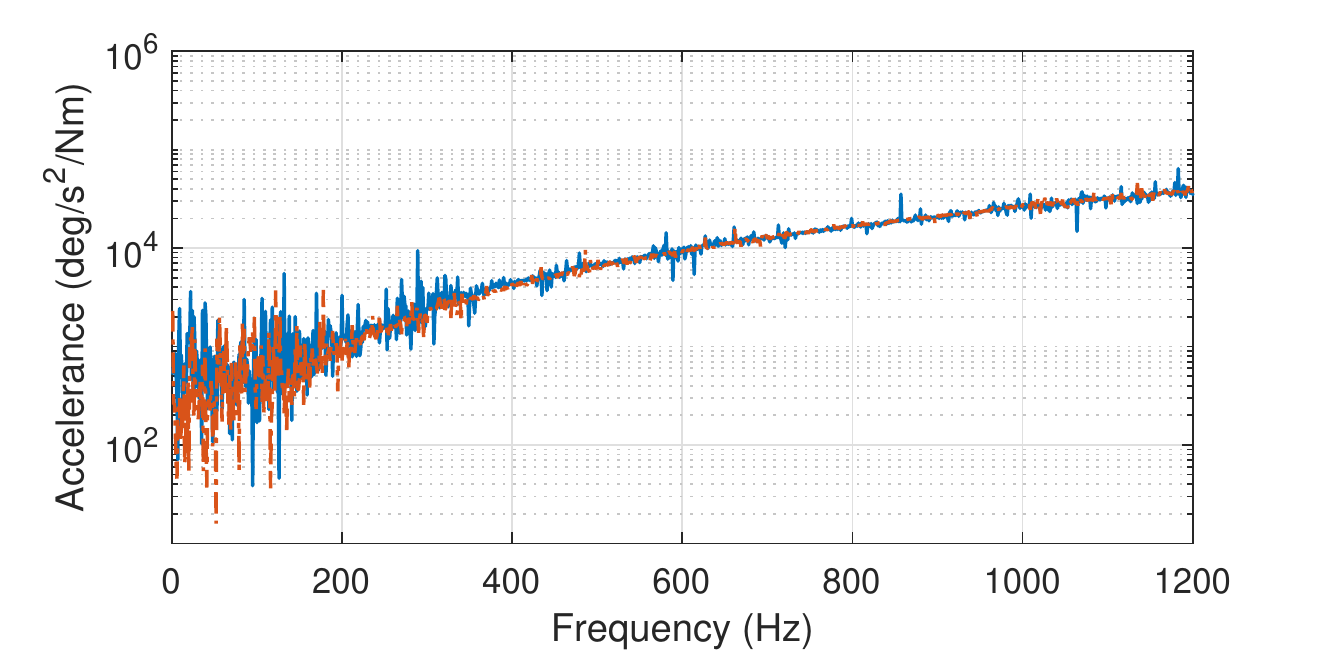}
		\caption{$\mathbf{Y}^{J}_{22,22}$ RZ/RZ}
		\label{fig:the_joint_rot_good}
	\end{subfigure}
	\caption{The decoupled joint accelerance on two translational DoF (a) and (b) and two rotational DoF (c) and (d) for both standard SEMM and correlated SEMM.}
	\label{fig:the_joint}
\end{figure}

The joint is decoupled using the methodology described in Section~\ref{sec:coupled_models} with the substructure hybrid models computed in Section~\ref{sec:filtering_application}. The joint accelerance is obtained when the expansion error $\epsilon$ defined in Section~\ref{sec:joint_decoupling} does not change anymore \cite{Klaassen2019}. The joint is decoupled (or identified) by finding its accelerance using the two methods:
\begin{itemize}
	\item Standard SEMM: $\quad  \mathbf{Y}^{J} = fbs(\mathbf{Y},\mathbf{B}) \quad \text{with} \quad \mathbf{Y} = diag(\mathbf{Y}^{\text{S},AJB}, -\mathbf{Y}^{\text{S},A}, -\mathbf{Y}^{\text{S},B})$
	\item Correlated SEMM: $\quad  \hat{\mathbf{Y}}^{J} = fbs(\hat{\mathbf{Y}},\mathbf{B}) \quad \text{with} \quad \hat{\mathbf{Y}} = diag(\hat{\mathbf{Y}}^{\text{S},AJB}, -\hat{\mathbf{Y}}^{\text{S},A}, -\hat{\mathbf{Y}}^{\text{S},B})$
\end{itemize}
Since accelerations are measured on all the structures with their FRFs expressed as accelerance, the identified joint is also represented as accelerance in Fig.~\ref{fig:the_joint}. As explained in Section~\ref{sec:interface_vp}, the joint is represented by two virtual points (12 DoF on each component) for the whole blade-disk interface i.e.\ a $24 \times 24$ system. 
In the past, the identified joints were limited to small systems under simplified motion and assumptions. To note a few examples for rigid joints, a $2 \times 2$ spring-damper system  was identified in \cite{Tol2015}, and a $4 \times 4$ spring-mass-damper system in \cite{Ren1995}. For our realistic case, the complexity of the motion of the interface requires a larger system description with the size of $24 \times 24$ -- in the form of accelerance FRFs \cite{Saeed2019,Saeed2020b}. 

Amongst those FRFs, only four accelerance (amplitude) plots are discussed as representative of the identified joint for both the standard and correlated SEMM in Fig.~\ref{fig:the_joint}. The first two plots are for the translational DoF (Fig.~\ref{fig:the_joint_trans_bad} and Fig.~\ref{fig:the_joint_trans_good}), and the other two are for the rotational DoF (Fig.~\ref{fig:the_joint_rot_bad} and Fig.~\ref{fig:the_joint_rot_good}). It is seen in the figures that despite some noisy behaviour, which is typical after a decoupling procedure \cite{Tsai1988,Ren1998,Wang2004,Tol2015}, the joint seems to follow a trend. 
This behaviour is representative of a system with high stiffness, low damping and low mass i.e.\ a stiffness dominant line on a logarithmic scale \cite{Ewins1995}. 
The fluctuations are due to the measurement and modelling errors, which propagate in the hybrid models. It can be observed that the accelerances obtained by the correlated SEMM exhibit slightly less fluctuations than the ones obtained by standard SEMM. Correlated SEMM in fact removes the channels that introduce more variability in the identification. However, the fluctuations still remain because the measurement noise cannot be completely removed.


It can be noticed in the translational accelerance (Fig.~\ref{fig:the_joint_trans_good}) and in the rotational one (Fig.~\ref{fig:the_joint_rot_bad}) that around 100 Hz, the accelerances from standard SEMM have a kind of a hump. This could be interpreted as an internal resonance of the joint. However, the hump disappears in the corresponding accelerance identified by the correlated SEMM, confirming that it was due to some spurious, non-physical effects which were eliminated by the correlated SEMM.

In the reconstruction of the response of the assembled system (blade plus disk with the joint in between), the accelerance curves of Fig.~\ref{fig:the_joint} were kept as they are (with their  fluctuations) without any fitting. This is due to the fact that the joint system is large and has high fluctuations (Fig.~\ref{fig:the_joint_trans_bad}). It would require curve-fitting to every FRF (a total of $576 \times 2$ curves including real and imaginary parts) in the joint accelerance with fluctuations while maintaining good matrix conditioning and symmetry. It is in itself a challenging task and is considered beyond the scope of this paper.


%
\begin{figure}[t!]
	\centering
	\includegraphics[scale=0.7]{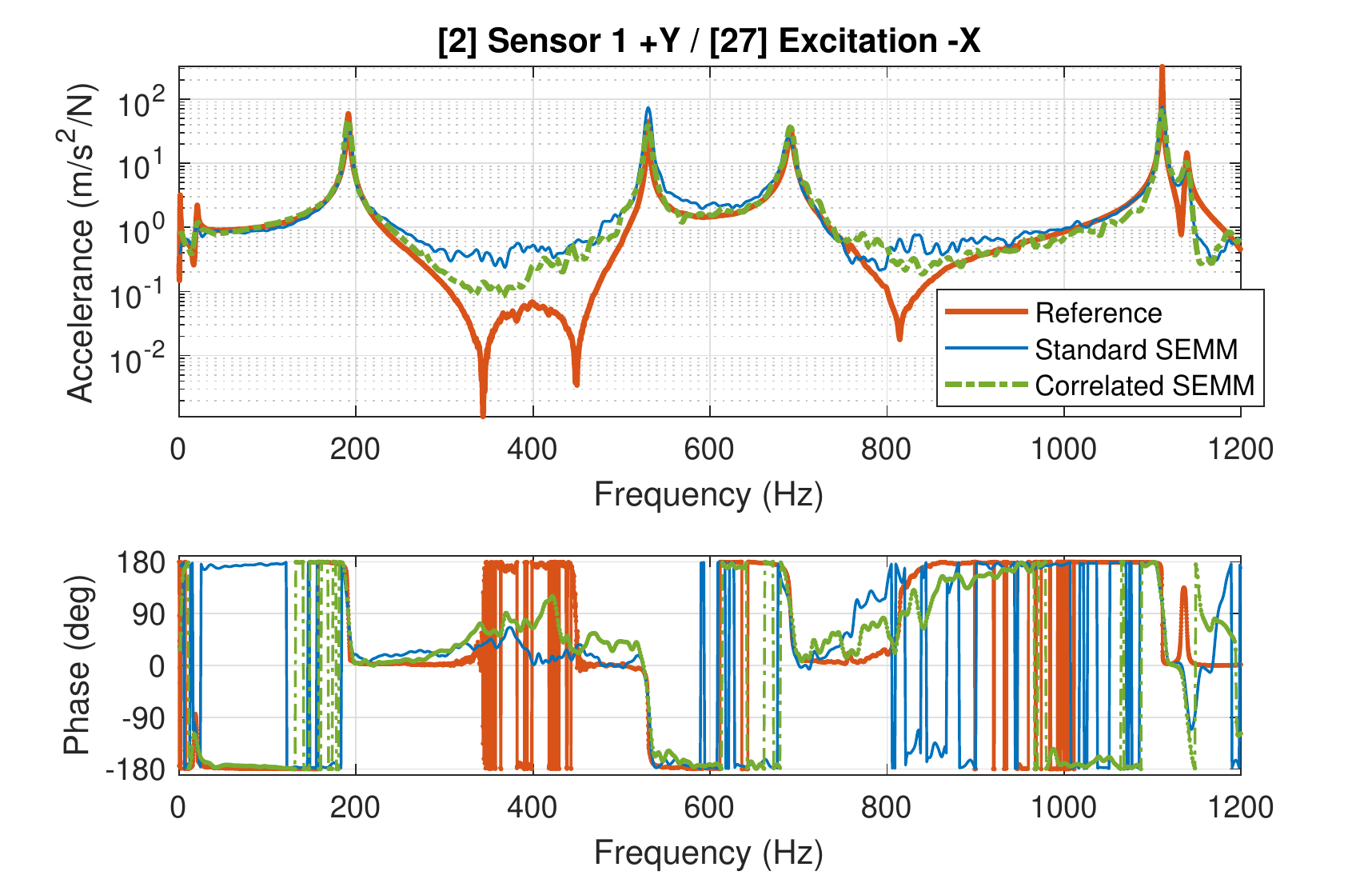}
	\caption{FRF on the coupled blade and assembly. The reference measured FRF is $\mathbf{Y}^{\text{exp},AB}_{2,27}$ (see Fig.~\ref{fig:sensors_impacts}). The FRFs with standard SEMM and correlated SEMM are obtained by recoupling the identified joint with their respective component hybrid models. The FRFs have been smoothed for clarity.}
	\label{fig:coupled_FRF}
\end{figure}

\subsection{Validation}
In order to check the reliability of the two identified joints, the identified joints are recoupled to their respective substructure models i.e.\ $\mathbf{Y}^{J}$ is recoupled to $\mathbf{Y}^{\text{S},A}, \mathbf{Y}^{\text{S},B}$ for the standard SEMM and $\hat{\mathbf{Y}}^{J}$ is recoupled to $\hat{\mathbf{Y}}^{\text{S},A} ,\hat{\mathbf{Y}}^{\text{S},B}$ for the correlated SEMM. 
Note that the joint accelerance of Fig.~\ref{fig:the_joint} is coupled as such without any fitting to the respective substructures due to the above-cited difficulties. 
A similar approach can be found in other works in literature \cite{DAmbrogio2014a,Sturm2017,Haeussler2020a}. In fact, the authors in \cite{DAmbrogio2014a} during identification of their known mass by substructure decoupling method noted that it was quite straightforward to detect errors in case of the known mass; however, it would not be the case if the system to be identified is unknown. They went on to assert that the only check that can be performed is to couple the predicted FRFs of the unknown subsystem with those of the known subsystem (to form the mathematically coupled system) and to compare it with the reference measured FRF on the assembly.
This type of validation is called on-board validation \cite{Sturm2017,Haeussler2020a} i.e.\ the reference FRFs (not included in the identification) in the same measurement campaign should be predicted by recoupling of the identified joint with the respective substructure models.

The recoupled accelerance FRFs obtained with both correlated and standard SEMM are shown in Fig.~\ref{fig:coupled_FRF} together with the FRF assumed as reference. The reference FRF was kept for validation since it was not included in the SEMM expansion. The values of the amplitude of the FRF peaks obtained for both standard and correlated SEMM are also listed in Table~\ref{tab:resonance_amplitudes} for the sake of comparison with the reference ones.
By looking just at the FRF of Fig.~\ref{fig:coupled_FRF}, it can be noticed that both the recoupled FRFs (standard and correlated) are almost overlapped to the reference curve in the regions close to the peak resonances. 
It can also be observed that, using the correlated  SEMM, leads to a general improvement in the regions of small amplitudes (ranges 220 - 740 Hz, 300 – 500 Hz, 800 - 900 Hz) where the FRF estimated by correlated SEMM is closer (than the one obtained by standard SEMM) to the reference FRF.

By looking in detail at the values of the peaks' amplitude in Table~\ref{tab:resonance_amplitudes}, the reader may observe that the values predicted by the correlated SEMM at resonance are in general better than the standard SEMM (except for mode 4 where the difference is negligible). In particular, the amplitude values predicted by correlated SEMM are much better for the first two peaks i.e.\ below 600 Hz.  This improvement is given by the correlated SEMM, instead of standard, in the model of the disk, and this was particularly effective for the disk in 0-600 Hz range (see Fig.~\ref{fig:FRF_disk_filtered}). From Table~\ref{tab:resonance_amplitudes}, it can be seen that, using the correlated SEMM, it is still advantageous in high frequency modes (3 to 5) even if the difference with standard SEMM is not always as evident as for modes 1 and 2. In these high frequency regions, other factors such as interface definition and singular value filtering, as shown in \cite{Saeed2020b} can also play a key role in better predicting the coupled system's dynamics.

\begin{table}[t]
	\centering
	\resizebox{\textwidth}{!}{%
		\begin{tabular}{ccc|cc|cc}
			\textbf{Mode} & \textbf{Frequency} & \textbf{Experiment} & \multicolumn{2}{c}{\textbf{Standard SEMM}} & \multicolumn{2}{c}{\textbf{Correlated SEMM}} \\ 
			& (Hz) & Amplitude & Amplitude & \% Difference & Amplitude & \% Difference \\ \hline 
			1 &	191.3 &	59.3 &	48.6 & -18.0\%	& 60.3 & 1.7\% \\
			2 & 530.0 &	44.7 &	91.4 &	104.6\% & 48.5 & 8.5\% \\
			3 & 691.1 &	33.6 &	25.5 & -24.2\%	& 41.5 & 23.3\% \\
			4 & 1111.1 & 298.3	& 99.0 & -66.8\% & 89.7	& -69.9\% \\
			5 & 1138.9 & 14.4 &	8.3	& -42.7\% &	12.2 &	-15.5\% \\
			\hline		
	\end{tabular}}
	\caption{Peak value comparison for the FRFs reconstructed by the standard SEMM and correlated SEMM methods. All amplitudes are in m/s$^2$/N.}
	\label{tab:resonance_amplitudes}
\end{table}
%
%
\section{CONCLUSIONS}
In this paper, a procedure to identify a dove-tail joint between a blade and disk is presented. The identification is based on a substructure decoupling technique. The novelty of the paper is that a new correlation based method is proposed in order to better select the experimental results to include in the joint identification procedure.

Considering that in a typical dove-tail joint, it is not possible to measure directly on the interfaces, the dynamics at the interface are then predicted by measurements at accessible points far from the joint. The already existing technique of System Equivalent Model Mixing (SEMM) is used here to generate hybrid numerical-experimental models of the single components (blade and disk) and of the blade-disk assembly. The hybrid models of each component (blade or disk) allow to predict the dynamics at the interface, that are coupled by the joint, by measuring in points far from the interface. However, the accuracy of these hybrid models is affected by two main error sources: i) expansion error (systematic or bias error) and ii) measurement errors (random as well as bias). The expansion error depends on the difference between the numerical and overlay (experimental) models. A big source of this error comes from the boundary conditions, for example, bad modelling of the component constraint. 

Therefore, in order to reduce at minimum the expansion error, it was here chosen to model both the structural components (blade and disk) in free conditions. As a result, a very good agreement between experimental and expanded FRFs from SEMM was obtained for each of the two components (the blade and disk).

The second source of error – the measurement errors – introduce noise and local inconsistencies in the FRFs. To reduce the effect of these errors, a procedure to check the goodness and usefulness of the measurements is here proposed. The procedure employs the FRAC (Frequency Response Assurance Criteria) correlation based method. 

When this procedure is introduced in SEMM it allows to systematically identify the poorly correlated measurement channels (that is the measurements polluting the construction of the hybrid model). This new improved approach (called correlated SEMM) here developed has the ability of filtering out the bad measurements. The correlated SEMM produces multiple hybrid models and each time computes correlations with a measured FRF taken as reference. The measurements with inconsistencies can be identified due to their low correlations levels and can be filtered out. This improved the overall quality of hybrid models of the two substructures (blade and disk).

The two SEMM methods (both standard and correlated) were here applied to the blade-disk assembly to decouple (identify) the joint. 
The result of the identification is the accelerance for each joint DoF. To validate this joint identification, the obtained joint accelerances were coupled back to the two hybrid models of the substructures to obtain the FRF of the assembled structure (blade plus disk). The obtained recoupled FRFs were then compared with the experimental FRFs measured on the assembled structure. 

This validation procedure was implemented both by using the standard SEMM and the correlated SEMM approaches. 
The recoupled FRF obtained by the correlated SEMM proved to be much more overlapped to the measured reference FRF than the FRF obtained by standard SEMM. In particular, the correlated SEMM showed to capture better the FRF plot in the non-resonance ranges.

The procedure implemented in the correlated SEMM of filtering out the low correlated measurements proved then to be effective also in the low response (near anti-resonance) regions which are more influenced by the noise level and the measurement errors.

\section*{ACKNOWLEDGMENTS}
This work is a part of the project EXPERTISE that received funding from the European Union's H2020 research and innovation program under the Marie Sk\l odowska-Curie grant agreement No 721865. The authors also acknowledge the efforts of Meysam Kazeminasab for conducting the experimental campaigns. 

\bibliography{referencesMendeley}

\end{document}